\documentclass[10pt]{article}

\usepackage{amsmath}
\usepackage{amsfonts}
\usepackage{amssymb}
\usepackage{graphicx}
\usepackage{setspace}
\usepackage{authblk}
\usepackage{cite}
\usepackage{setspace}

\usepackage[linktocpage=true]{hyperref}
\usepackage{float}
\usepackage[dvipsnames]{xcolor}
\usepackage{hyperref}
\hypersetup{colorlinks,linkcolor={blue},citecolor={red},urlcolor={blue},filecolor={Turquoise}}

\begin{document}
	
	\title{\textbf{Phenomenology of an $E_6$ inspired extension of Standard Model: Higgs sector} }
	\author{Sanchari Bhattacharyya\footnote{sanchari1192@gmail.com}\;}
	\author{\;Anindya Datta\footnote{adphys@caluniv.ac.in}}
	\affil{\emph{Department of Physics, University of Calcutta} \\ \emph{92, Acharya Prafulla Chandra Road, Kolkata 700009}}
	\date{}
	\maketitle
	
	\vskip 2cm
	
	\begin{abstract}
			We investigate a variant of Left-Right symmetric model based  $SU(3)_C \otimes SU(2)_L \otimes U(1)_L \otimes SU(2)_R \otimes U(1)_R$ gauge group $(32121)$. Spontaneous breaking of $32121$ down to the Standard Model (SM) gauge group, requires a bi-doublet under $SU(2)_L \otimes SU(2)_R$, a right-handed doublet scalar under $SU(2)_R$  along with a $SU(2)$ singlet scalar boson. Symmetry breaking leads to several  neutral and charged massive gauge bosons apart from the SM $W$ and $Z$. The Large Hadron Collider (LHC) results for the search of heavy gauge bosons can be used to constrain the vacuum expectation values responsible for giving masses to these extra heavy gauge bosons. Physical spectrum of the scalar bosons contains several neutral CP-even and CP-odd states and a couple of charged scalars apart from the SM-like Higgs boson. We have put constraints on the masses of some of these scalars from the existing LHC data. The possible decay rates and production cross-sections of these scalars have been investigated in some details. Production cross-sections for some of the scalars look promising at 14 TeV and 27 TeV run of the LHC with high luminosity option. We keep, in our model,  all the fermions present in the $27$-dimensional fundamental representation of $E_6$. Mass limit of one such  exotic lepton has also been derived from present LHC data. It is noted that some of these neutral exotic lepton or neutral scalar bosons  of this model can serve the purpose of a cold dark matter. 
			
	\end{abstract}
		
		
	\hrulefill

	\section{Introduction}
	The Standard Model (SM) of Particle Physics has been extremely successful in describing the interactions of elementary particles and fundamental forces operative in microscopic world. Probably, the most subtle prediction of the SM has been  the existence of a scalar boson, the Higgs boson, responsible for giving masses to all the elementary particles. With the discovery of Higgs boson \cite{higgs_atlas, higgs_cms} at the Large Hadron Collider (LHC), CERN;  all the predictions of the SM has been tested. In spite of its immense success, SM in its original form, miserably fails to explain one  important piece of experimental observations, namely, the existence of Dark Matter (DM), a new kind of very weakly interacting but massive matter pervading the whole universe. Moreover, it is very crucial to know whether the discovered 125 GeV Higgs boson is the sole agent for Electroweak Symmetry Breaking (EWSB) or a more rich scalar sector is responsible for such an act.  There are several theoretical studies  \cite{Higgs-review} which have  been devoted to investigate the phenomenology of extended Higgs sectors. It is important to note that any model with extended scalar sector must contain  a physical CP-even scalar boson with exactly the same properties of the SM Higgs. 
	The extended scalar sector may also  be instrumental in resolving some of the shortcomings of the SM. As for example, extended Higgs sectors with singlet scalars may resolve the problem of dark matter \cite{singlet-DM}. Left-Right (LR) symmetric triplet Higgs models are very popular in explaining neutrino masses via  seesaw mechanisms \cite{triplet-neutrinomass}, \cite{Nu_mass1}. Multi-Higgs doublet models have been used in explaining flavour problems \cite{2hdm}. An added bonus for many such models with extended Higgs sector is possibility of a stable neutral scalar which may act as a suitable candidate for DM. On the experimental front, signatures for extra neutral and charged scalar bosons have already been in the top of the agenda for all the past and present experimental programmes. Unfortunately, the evidence for the SM-like Higgs boson has been observed so far. Present precision of experimental data on Higgs signal strengths in different channels and our lack of experimental knowledge on Higgs tri-linear coupling limit us from conclusively decide whether the 125 GeV boson is the only agent of EWSB \cite{higgs-precision}. In near future, with high luminosity (HL) 14 (27) TeV run of the LHC we expect these pictures to be more clearer. So it is of utmost priority for particle physics community to construct such models with extended scalar and/or gauge sector and check whether these models are consistent with the available and future experimental data from the LHC. A number of such models have been proposed and their possible experimental signatures at the LHC have been sought for. Unfortunately, many such efforts have been futile so far. Non-observation of any signature of Physics beyond the SM (whether supersymmetry \cite{SUSY-search} or extra-dimension \cite{ed-search}), only pushes the scale of such new dynamics in upward direction.	

	In an endeavour to construct a model which has a rich scalar sector satisfying the LHC data  we turn our attention to an unifying gauge group $E_6$\cite{E6}, which can be broken down to $\left[ SU(3) \right]^3$ followed by a further breakdown to $SU(3)_C \otimes SU(2)_L \otimes U(1)_L \otimes SU(2)_R \otimes U(1)_R$ ($32121$). We will not be interested in the dynamics which may be responsible for the aforementioned symmetry breaking chain, rather we will be investigating the appearance of SM gauge group from $32121$ and the resulting phenomenology of additional particles and their interactions among themselves or with the particles of SM. The main advantage of working in a framework of unifying group like $E_6$  is the natural appearance of right-handed neutrinos as well as 3 generations of heavy neutral leptons, singlet under either of $SU(2)$ groups. The right-handed heavy neutrinos may be responsible for neutrino mass generations via the (Type-I) seesaw mechanism. This model contains a large number of neutral and charged scalar bosons after the spontaneous breakdown of $SU(2)_L \otimes U(1)_L \otimes SU(2)_R \otimes U(1)_R$ to EW gauge group of the SM. These color singlet scalars originate from (\textbf{1}, \textbf{3}, $\bar{\textbf{3}}$) representation of  $[SU(3)]^3$. The presence of such rich scalar sector is one of the reasons behind the present investigation. Furthermore, several exotic fermions will arise from the $27$-dimensional fundamental representation of $E_6$. Some of the neutral Higgs bosons and charge neutral leptons may also serve the purpose of dark matter. However, in this article we will mainly investigate the phenomenology of the extended scalar sector. A separate study will be devoted to the DM aspect of this model.

Although, the scalar sector of the model under consideration have many similarities with that of the left-right model (LRM), the phenomenology of the 
scalar sector has some distinct features which are different from left-right symmetric model (LRSM). In the present work, we have banked on these features 
of the model to distinguish it from the often studied models like LRSM.  Let us emphasise on the novelty of the present work in 
the following.

\begin{itemize}
\item Although the $32121$ gauge group respects the LR symmetry, the model under consideration is different from the conventional LRS model. The 
first hint of this difference comes from the fact that both the bi-doublet vacuum expectation values (vevs) $k_1$ and $k_2$ cannot be simultaneously set to non-zero values unlike 
its more familiar variant based on $SU(2)_L \times SU(2)_R \times U(1)_{B-L}$
unless we consider a tri-linear term in the scalar potential.  Even after adding such a term which comes out to be small along with 
small value of $k_2$, phenomenology of this model remains practically the same as $k_2 =0$ case. This makes the phenomenology of the scalar sector of $32121$ model different from LRSM. 

\item We have considered the complete set of fermions arising from the $\bf 27$-dimensional representation of $E_6$. Some of these
 fermions  are stable but have electric charges. We have for the first time derived bounds on their masses 
 from the present LHC data.
 A detailed endeavour from the experimentalist friends 
 to study their signature including detector simulation is presently unavailable and is urgently needed. 
 
 \item Although one of the charged Higgs bosons arising in the model has conventional decay to $tb$, the other one is stable 
 and produces a charged track in the detector. Such stable charged bosons are not present in the LRSM and their signatures have not 
 been discussed in the present literature so far. 
 
 \item The pair-production of charged and neutral Higgs bosons arising from left-handed doublet will produce either two charged tracks or a single charged track in the detector, giving rise to a background free novel signature of this model.  
 
 \item Presence of a heavy neutral gauge boson,$A'$ apart from $Z'$ (arising because of the presence of $SU(2)_R$, can be identified with the $Z_R$ in LRSM) is also a hall mark of the model under the consideration. This 
 neutral boson arises due to the extra $U(1)$ factor in $32121$ and couples to all the SM fermions, in contrast to several extensions of the SM by an extra $U(1)$, where this heavy neutral state have selective coupling with the SM fermions. Unlike the LRS model, where only one heavy neutral gauge boson is present, $32121$ is characterised by two such states with similar properties but having different masses.
 
 \item Three possible Dark Matter candidates in forms of exotic fermion, a scalar and a pseudoscalar originating from left-doublet, are present in the spectrum. Although we will not discuss the direct detection bounds or relic density, an  lower bound on the mass of the scalar has been derived indirectly from the LHC data in the present analysis.

\end{itemize}	

Before we come to an end of this section, it is important to mention that we have only guided by the framework of $E_6$ to study the phenomenology of the 
a set of scalars whose masses are somehow interlaced due to the structure of the scalar potential respecting the local $32121$ gauge symmetry. Similarly,
the gauge group and complete set of $27$-plet of of fermions are chosen from a low energy point of view. Let us re-emphasize that we will not study the symmetry breaking chain $E_6 \rightarrow \left[ SU(3) \right]^3 \rightarrow 32121$, neither we start with an unifying Yukawa texture appropriate for $E_6$ starting from GUT scale to generate the same at EW scale by renormalization group equation (RGE) running. The Yukawa matrices in our analysis, have been assumed to be non-diagonal. However, we will not discuss the pattern of masses and mixings for the exotic fermions present in the spectrum. 
The physical masses and mixings (in case of SM sector) have been assumed to be in consistence with the observed values of SM fermions
while, for the exotic sector, they are free parameters defined at the EW scale. So, our connection to $E_6$ is only confined to the choice of 
gauge group and choice of fermions at low energy, without considering any effect from high scale physics creeping in due to renormalization.

	Plan of the article is  the following. In Section \ref{section2}, we discuss the symmetry breaking mechanism and spectra of Higgs bosons after Spontaneous Symmetry Breaking (SSB). This section will also contain a discussion of gauge boson masses and fermion Yukawa interaction with the scalars. In the next section we will discuss in detail the phenomenology of the physical Higgs bosons which arise in the model after SSB. The decay branching ratios (BR) and production cross-sections of such Higgs bosons are presented in Section \ref{section3} in the context of 14 and 27 TeV run of the LHC. Finally, we conclude in Section \ref{section4} .

	\section{The description of $SU(3)_C \otimes SU(2)_L \otimes U(1)_L \otimes SU(2)_R \otimes U(1)_R$ model}\label{section2}
	In this article, we will be interested in an extension of the SM  whose root can be traced back to the Grand Unified Group  $E_6$. However at the energy scale of LHC experiment in which we are interested, we keep all the fermions and Higgs multiplets excepting few colored scalar bosons, which naturally arise in the 27-dimensional fundamental representation of $E_6$. However, to keep the number of matter fields in our model to a minimum, we will assume that the colored scalars are too heavy (of the order of symmetry breaking scale of $\left[ SU(3) \right]^3$) to affect the phenomenology at the LHC energy\footnote{If we intend to break the $\left[ SU(3) \right]^3$  to $31221$ by embedding the former into a 5-dimensional manifold and applying appropriate orbifold boundary conditions, one may get rid of such colored scalars by choosing appropriate boundary conditions on these fields at the orbifold boundaries. \cite{orbifold}}.  
	Higgs multiplets will be instrumental in breaking  down $SU(3)_C \otimes SU(2)_L \otimes U(1)_L \otimes SU(2)_R \otimes U(1)_R$ to the SM the gauge group. While the complete set of fermions present in $\textbf{27}$-plet are necessary for anomaly cancellation.  
	However, this choice of fermion representations used in our analysis is no way unique. One can have vanishing anomaly contribution only by considering the left and right-handed fermion doublets $L_L$, $L_R$, $Q_L$ and $Q_R$. Hyper-charge assignments of the rest of the fermion multiplets listed in Table \ref{table1} can then be done by the consideration of anomaly cancellation among the exotic fermions and can be done in more than one ways. We have presented two such cases in the Table \ref{table2}.  It is important to note that in such a situation, $U(1)_{L,R}$ charges for the Higgs multiplets will be different from the case presented in this analysis. However, we will not consider such a situation and present our phenomenological analysis with complete set of fermions arising from the $27$-plet.

	Gauge bosons present in this model automatically follows from the gauge group of our interest. The matter and gauge fields which are present in our model are listed in Table \ref{table1}\footnote{Electric charge, $Q$ is defined through the relation, $Q = T_{3L} + T_{3R} + Y_L/2 + Y_R/2$. $L$ and $R$ carry their usual meaning. These quantum numbers for each multiplet of $32121$ model have been noted in the Table \ref{table1}}.

	\begin{table}[h!]
		\centering
		\begin{tabular}{|c|c|c|c|c|c|c|}
			\hline \hline
			&  & $3_C$ & $2_L$ & $2_R$ & $1_L$ & $1_R$ \\ [1.0ex]
			\hline 
%
			& $L_L$ & $1$ & $2$ & $1$ & $-1/6$ & $-1/3$  \\ [1.0ex]
			& $\bar{L}_R$ & $1$ & $1$ & $2$ & $1/3$ & $1/6$  \\ [1.0ex]
			& $\bar{L}_B$ & $1$ & $2$ & $2$ & $-1/6$ & $1/6$  \\ [1.0ex]
			Fermions & $\bar{L}_S$ & $1$ & $1$ & $1$ & $1/3$ & $-1/3$ \\ [1.0ex]
			& $Q_L$ & $3$ & $2$ & $1$ & $1/6$ & $0$ \\ [1.0ex]
			& $\bar{Q}_R$ & $\bar{3}$ & $1$ & $2$ & $0$ & $-1/6$ \\ [1.0ex]
			& $\bar{Q}_{LS}$ & $\bar{3}$ & $1$ & $1$ & $-1/3$ & $0$ \\ [1.0ex]
			& $Q_{RS}$ & $3$ & $1$ & $1$ & $0$ & $1/3$ \\ [1.5ex]
			\hline
			& $\Phi_B$ & $1$ & $2$ & $2$ & $1/6$ & $-1/6$  \\ [1.0ex]
			Bosons & $\Phi_L$ & $1$ & $2$ & $1$ & $1/6$ & $1/3$  \\ [1.0ex]
			& $\Phi_R$ & $1$ & $1$ & $2$ & $-1/3$ & $-1/6$  \\ [1.0ex]
			& $\Phi_S$ & $1$ & $1$ & $1$ & $-1/3$ & $1/3$ \\ [1.5ex]
			\hline	
			& $G^i ,\; i=1,...,8$& $8$ & $1$ & $1$ & $0$ & $0$ \\ [1.0ex]
			& $W^i_{L}, i=1,2,3$ & $1$ & $3$ & $1$ & $0$ & $0$ \\ [1.0ex]
			Gauge bosons & $W^i_{R}, i=1,2,3$ & $1$ & $1$ & $3$ & $0$ & $0$  \\ [1.0ex]
			& $B_L$ & $1$ & $1$ & $1$ & $0$ & $0$ \\ [1.0ex]
			& $B_R$ & $1$ & $1$ & $1$ & $0$ & $0$ \\ 
			\hline
		\end{tabular}
		\caption{Fermions and Bosons in  $32121$ model with their respective quantum numbers}
		\label{table1}
	\end{table}

	\begin{table}[h!]
		\centering
		\begin{tabular}{|c|c|c|c|c|c|}
			\hline \hline
			& $3_C$ & $2_L$ & $2_R$ & $1_L$ & $1_R$ \\ [1.0ex]
			\hline 
			$L_L$ & $1$ & $2$ & $1$ & $1$ & $-3/2$  \\ [1.0ex]
			$\bar{L}_R$ & $1$ & $1$ & $2$ & $-1$ & $3/2$  \\ [1.0ex]
			$Q_L$ & $3$ & $2$ & $1$ & $-1/3$ & $1/2$ \\ [1.0ex]
			$\bar{Q}_R$ & $\bar{3}$ & $1$ & $2$ & $1/3$ & $-1/2$ \\ [1.5ex]
			\hline				
			$\Phi_B$ & $1$ & $2$ & $2$ & $0$ & $0$  \\ [1.0ex]
			$\Phi_R$ & $1$ & $1$ & $2$ & $1$ & $-3/2$  \\ [1.0ex]
			$\Phi_S$ & $1$ & $1$ & $1$ & $0$ & $0$ \\
			\hline \hline
			$\bar{L}_B$ & $1$ & $2$ & $2$ & $0$ & $0$  \\ [1.0ex]
			$\bar{Q}_{LS}$ & $\bar{3}$ & $1$ & $1$ & $1/3$ & $1/3$ \\ [1.0ex]
			$Q_{RS}$ & $3$ & $1$ & $1$ & $-1/3$ & $-1/3$ \\ [1.0ex]
			$\bar{L}_S$ & $1$ & $1$ & $1$ & $0$ & $0$ \\
			\hline\hline
		\end{tabular}
		\caption{Fermions and Scalars in $32121$ model with their respective quantum numbers. The important thing to note here that, the $U(1)_{L}$ hypercharges of $Q_{LS}$ and $Q_{RS}$ have to be same to cancel chiral anomaly. Similar for the case of $U(1)_{R}$ too. But $U(1)_{L}$ and $U(1)_{R}$ hyperchages for each of them may not be equal. That depends upon the charge assignment of $Q_{LS},Q_{RS}$. However, we consider $Q_{LS},Q_{RS}$ have $2/3$ charge. So it is an easy way to choose the $U(1)_{L}$ and $U(1)_{R}$ hypercharges of $Q_{LS}$ as well as $Q_{RS}$ to be equal i.e., $1/3$.}
		\label{table2}
	\end{table}	
	
	The first block represents the minimal matter contents in the fermion sector, i.e., SM-fermions with right-handed neutrino. The scalars in the next block will provide us a gauge invariant Yukawa Lagrangian generating the masses of the SM-fermions and Majorana mass for $\nu_R$. We can generate the masses of all the exotic fermions with $\Phi_S$. 
	
	The third block represents the other beyond Standard Model (BSM) fermions present in \textbf{27}-plet and separately cancels chiral anomaly. $L_B$ is SU(2) doublet but U(1) singlet, $L_S$ is pure singlet. Here are two cases. For $Q_{LS}$ as well as $Q_{RS}$ separately,
	
	$(i)$ If $U(1)_{L}$ hypercharge is equal to $U(1)_{R}$ hypercharge (i.e., same value/sign), $Q_{LS},Q_{RS}$ will be vector-like and they will not contribute in anomaly cancellation.
	
	$(ii)$ If $U(1)_{L}$ hypercharge is not equal to $U(1)_{R}$ hypercharge (i.e., different value/sign), $Q_{LS},Q_{RS}$ will be chiral fermions and will contribute in anomaly cancellation.

	\subsection{Scalar sector of $SU(3)_C \otimes SU(2)_L \otimes U(1)_L \otimes SU(2)_R \otimes U(1)_R$ model}
	We now briefly discuss the Higgs multiplets responsible for symmetry breaking and their interactions in this model. 
	The scalar sector of the $32121$ model contains one Higgs bi-doublet ($\Phi_B$), one left-handed ($\Phi_L$), one right-handed ($\Phi_R$) weak doublets and a singlet Higgs boson ($\Phi_S$) with non-zero $U(1)$ charges. These scalars arise from the (\textbf{1}, \textbf{3}, $\bar{\textbf{3}}$) representation of $\left[ SU(3)\right]^3$. For a complete symmetry breaking mechanism from $32121 \longrightarrow SU(3)_C \otimes SU(2)_L \otimes U(1)_Y \longrightarrow SU(3)_C \otimes U(1)_{EM}$, the alignments of  Higgs bi-doublet, right (left)-handed doublet and the singlet will be the following.

	\begin{eqnarray}
	\Phi_B &=& \begin{pmatrix}
	\frac{1}{\sqrt{2}}(k_1 + h_1^0 + i \xi_1^0)  & h_1^+  \\ h_2^- & \frac{1}{\sqrt{2}}(k_2 + h_2^0 + i \xi_2^0)
	\end{pmatrix},  \nonumber \\
	\vspace{1cm}
	\Phi_L &=& \begin{pmatrix}
	h_L^+ \\ \frac{1}{\sqrt{2}}(v_L + h_L^0 + i \xi_L^0)
	\end{pmatrix}, 
	\Phi_R = \begin{pmatrix}
	\frac{1}{\sqrt{2}}(v_R + h_R^0 + i \xi_R^0) \\ h_R^- 
	\end{pmatrix}, \Phi_S = \frac{1}{\sqrt{2}}(v_S + h_S^0 + i \xi_S^0)
	\end{eqnarray}

	The Higgs potential, $\cal V$ which obeys symmetry of the gauge group can be written as sum of two parts 
		${\cal V}_1$ and ${\cal V}_2$, where,
	
	\begin{eqnarray}
	\label{pot}
	\mathcal{V}_1 = &-&\mu_1^2 Tr \left( {\Phi_B}^{\dagger} \Phi_B\right) - \mu_3^2 \left( {\Phi_L}^{\dagger} \Phi_L + {\Phi_R}^{\dagger} \Phi_R \right)  - \mu_4^2 {\Phi_S}^{\dagger} \Phi_S   \nonumber \\
	&+& \lambda_1 Tr \left[ ({\Phi_B}^{\dagger} \Phi_B)^2\right] + \lambda_3 \left( Tr\left[ {\Phi_B}^{\dagger} \tilde{\Phi}_B\right]  Tr\left[ \tilde{\Phi}_B^{\dagger} \Phi_B\right] \right) \nonumber \\
	&+& \alpha_1 (\Phi_S^{\dagger} \Phi_S)^2 + \beta_1 Tr\left[ {\Phi_B}^{\dagger} \Phi_B\right]  (\Phi_S^{\dagger} \Phi_S) + \gamma_1 \left[ (\Phi_L^{\dagger} \Phi_L) + (\Phi_R^{\dagger} \Phi_R)\right] (\Phi_S^{\dagger} \Phi_S) \nonumber \\
	&+& \rho_1 \left[ (\Phi_L^{\dagger} \Phi_L)^2 + (\Phi_R^{\dagger} \Phi_R)^2\right] 
	+ \rho_3 \left[ (\Phi_L^{\dagger} \Phi_L) (\Phi_R^{\dagger} \Phi_R)\right] + c_1 Tr\left[ {\Phi_B}^{\dagger} \Phi_B\right] \left[ (\Phi_L^{\dagger} \Phi_L) + (\Phi_R^{\dagger} \Phi_R)\right]   \nonumber \\
	&+& c_3 \left[  ( \Phi_L^{\dagger} \Phi_B  \Phi_B^{\dagger} \Phi_L ) + ( \Phi_R^{\dagger} \Phi_B^{\dagger} \Phi_B  \Phi_R ) \right] + c_4 \left[ ( \Phi_L^{\dagger} \tilde{\Phi}_B \tilde{\Phi}_B^{\dagger} \Phi_L ) + ( \Phi_R^{\dagger} \tilde{\Phi}_B^{\dagger} \tilde{\Phi}_B \Phi_R ) \right]
	\end{eqnarray}
	and, \begin{equation}
	\label{tri-linear}
	\mathcal{V}_2 = \mu_{BS} Tr \left[ {\Phi^\dagger _B} \tilde \Phi_B\right]  \Phi_S^\ast + h.c.
	\end{equation}	
	
	All the parameters in $\cal V$ are considered to be real excluding any possibility of CP-violation via the Higgs sector. 
	$\mathcal{V}$ has a symmetry under $L\longleftrightarrow R$  exchange. In the above,  $\tilde{\Phi}_B \equiv \sigma_2 \Phi_B^* \sigma_{2}$.

	We note that ${\cal V}_1$ has a symmetry corresponding to global phase transformations on the fields 
	\begin{equation}
	\Phi_B \rightarrow e^{i \theta_B}\; \Phi_B; ~~\Phi_L \rightarrow e^{i \theta_L}\; \Phi_L; ~~\Phi_R \rightarrow e^{i \theta_R}\; \Phi_R {~\rm and} 
	~\Phi_S \rightarrow e^{i \theta_S}\; \Phi_S.
	\label{global}
	\end{equation}
	
	However, ${\cal V}_2$ which is  proportional to $\mu_{BS}$, breaks this global symmetry explicitly (for example see ref. \cite{global_explicit}) otherwise respecting the symmetries of $32121$ gauge group. 	
	This results into appearance of bilinear terms like $h^0 _1 h^0 _2 $, $h^+ _1 h^- _2$.  However,  with both $k_1, k_2 \neq 0$ such bilinear terms also could be generated from the term proportional to $\lambda_3$. Setting one of these vevs to zero automatically prohibits the appearance of such bilinear terms in the scalar potential. In other words, setting both $k_1$ and $k_2$ to their non-zero values excluding ${\cal V}_2$, would break the global symmetry in Eq. \ref{global} spontaneously which results into undesirable extra massless scalar modes.  
	One can of course have non-zero $k_1$ and $k_2$ simultaneously, however in such a case, a tri-linear term proportional to $\mu_{BS}$ is necessary to break the global symmetry explicitly and thus avoiding the appearance of extra Goldstone modes.

	The kinetic Lagrangian for scalars is,
	\begin{equation}
	\label{LSK}
	\mathcal{L}_{\Phi} = Tr[{(\mathcal{D}_{\mu}\Phi_B)}^\dagger (\mathcal{D}^{\mu}\Phi_B)] + {(\mathcal{D}_{\mu}\Phi_L)}^\dagger (\mathcal{D}^{\mu}\Phi_L) + {(\mathcal{D}_{\mu}\Phi_R)}^\dagger (\mathcal{D}^{\mu}\Phi_R) + {(\mathcal{D}_{\mu}\Phi_S)}^\dagger (\mathcal{D}^{\mu}\Phi_S)
	\end{equation}
	
	It is needless to mention that covariant derivatives acting on different Higgs multiplets, are not same and contain appropriate gauge bosons in them.

		Minimization conditions we obtain are the following,
		\begin{eqnarray}
		\label{mu1k1}
		&& 2 \sqrt{2} \mu_{BS} k_2 v_S + k_1 (2 \lambda_1 k_1^2 + 2 (\lambda_1 + 2\lambda_3) k_2^2 - 2 \mu_1^2 + (c_1 +c_4) v_L^2 + (c_1+c_3) v_R^2 + \beta_1 v_S^2)= 0 \\
		\label{mu1k2}
		&& 2 \sqrt{2} \mu_{BS} k_1 v_S + k_2 (2 \lambda_1 k_2^2 + 2 (\lambda_1 + 2\lambda_3) k_1^2 - 2 \mu_1^2 + (c_1 +c_3) v_L^2 + (c_1+c_4) v_R^2 + \beta_1 v_S^2) = 0 \\
		\label{mu3vl}
		&& v_L \;[(c_1+c_4) k_1^2 + (c_1+c_3) k_2^2 - 2 \mu_3^2 + 2 \rho_1 v_L^2 + \rho_3 v_R^2 + \gamma_1 v_S^2] = 0 \\
		\label{mu3vr}
		&& v_R \;[(c_1+c_3) k_1^2 + (c_1+c_4) k_2^2 - 2 \mu_3^2 + 2 \rho_1 v_R^2 + \rho_3 v_L^2 + \gamma_1 v_S^2] = 0 \\
		\label{mu4vs}
		&& 2 \sqrt{2} k_1 k_2 \mu_{BS} + v_S (\beta_1 (k_1^2+k_2^2) - 2 \mu_4^2 + \gamma_1 (v_L^2+v_R^2) + 2 \alpha_1 v_S^2) = 0
		\end{eqnarray}
		
		From Eqs. \ref{mu1k1} and \ref{mu1k2},
		\begin{eqnarray}
		\label{mu1k1new}
		\mu_1^2 = \dfrac{1}{2 k_1} \left(2 \sqrt{2} \mu_{BS} k_2 v_S + k_1 (2 k_1^2 \lambda_1 + 2 (\lambda_1 + 2 \lambda_3) k_2^2 + (c_1 + c_4) v_L^2 + 
		(c_1 + c_3) v_R^2 +\beta_1 v_S^2) \right) \\
		\label{mu1k2new}
		\mu_1^2 = \dfrac{1}{2 k_2} \left(2 \sqrt{2} \mu_{BS} k_1 v_S + k_2 (2 k_2^2 \lambda_1 + 2 (\lambda_1 + 2 \lambda_3) k_1^2 + (c_1 + c_3) v_L^2 + 
		(c_1 + c_4) v_R^2 +\beta_1 v_S^2) \right)
		\end{eqnarray}
		Using Eqs. \ref{mu1k1new} and \ref{mu1k2new} for $k_1, k_2 \ne 0$, we have,
		\begin{equation}
		\label{mubs}
		\mu_{BS} = \dfrac{1}{\sqrt{2} v_S} \dfrac{k_1 k_2}{k_2^2-k_1^2} \left((c_3-c_4) \dfrac{v_L^2-v_R^2}{2} -2\lambda_3 (k_2^2-k_1^2) \right)
		\end{equation}
	
	

		Spontaneous breaking of Left-Right symmetry demands, $v_R \neq 0$.  Thus Eq. (\ref{mu3vr}) results into,
		\begin{equation}
		\label{mu3}
		\mu_3^2=\frac{1}{2}[(c_1+c_3)k_1^2+\rho_3 v_L^2+2 \rho_1 v_R^2+\gamma_1 v_S^2].
		\end{equation}
		The choice $v_L \neq 0$ leads to appearance of an extra massless scalar mode which is undesirable\footnote{This is related to the spontaneous breakdown of a global symmetry of the  defined in Eq. \ref{global}.} scalar potential. So we stick to the case with $v_L =0$.
		
		To break the extra $U(1)$ we have to opt for $v_S \neq 0$ resulting into (from Eq. \ref{mu4vs}),
		\begin{equation}
		\label{mu4}
		\mu_4^2 = \dfrac{1}{2 v_S} \left(2 \sqrt{2} k_1 k_2 \mu_{BS} + v_S (\beta_1 (k_1^2 + k_2^2) + \gamma_1 (v_L^2 + v_R^2) + 2 \alpha_1 v_S^2)\right)
		\end{equation}

	Once we fix the minimisation conditions of the Higgs potential, we are ready to note the Higgs mass matrices under such alignment of vacuum.  
	However before delving into the details of scalar mass matrices let us make some brief comments on an important issue related to the minimum of the scalar potential. It is important to note that a scalar potential such as Eq. \ref{pot} depending on so many fields may have more than one minima having varying depths. Thus merely satisfying the minimisation condition (by scalar potential parameters) does not ascertain that one is at the deepest minimum of the potential. In principle, different choices of the scalar potential parameters correspond to minima of different depths. Moreover, radiative corrections can significantly change to the structure of scalar potential and consequently change the depths of different minima of the potential. Hence, it is expected that one must at least incorporate one-loop corrections to the scalar  potential to before looking for the deepest minima. However the exercise of calculating an effective potential at one loop for our model is beyond the scope of the present analysis. So we stick to the tree level potential and have not tried to look for its deepest minima. As long as  the tunnelling time from the false vacuum to the true (deepest) vacuum is larger than the lifetime of the Universe, sitting at a minimum other than the deepest one is not always hazardous. However a realistic estimation of this tunnelling time also requires a one loop corrected effective potential of our model.  A recent study \cite{bhupaldev}  has been devoted to the analysis of the tree level scalar potential with particular emphasis on vacuum alignment and structure of minima of the potential. We would like to note that alignment of the vacuum used in our analysis satisfies the criterion of a {\em good vacuum} a la \cite{bhupaldev}.
	
	In the following, we note the CP-even, CP-odd and charged scalar mass matrices after replacing $\mu_1$, $\mu_{BS}$, $\mu_3$ and $\mu_4$ using Eqs. \ref{mu1k1new}, \ref{mubs}, \ref{mu3} and \ref{mu4} respectively.

		In a basis, defined by the fields $\left\lbrace h_1^0,h_2^0,h_L^0,h_R^0,h_S^0\right\rbrace $ the square of CP-even mass matrix (${M_r^0}^{2}$) is,
		\small
		\begin{eqnarray}
		\begin{pmatrix}
		2\lambda_1 k_1^2 + k_2^2 \Delta' & 2 \lambda_1 k_1 k_2 + k_1 k_2 \Delta' & 0 & (c_1+c_3) k_1 v_R & \beta_1 k_1 v_S -\dfrac{k_1 k_2^2 \Delta'}{v_S} \\
		2 \lambda_1 k_1 k_2 + k_1 k_2 \Delta' & 2\lambda_1 k_2^2 + k_1^2 \Delta' & 0 & (c_1+c_4) k_2 v_R &  \beta_1 k_2 v_S -\dfrac{k_1^2 k_2 \Delta'}{v_S} \\
		0 & 0 & \dfrac{1}{2} \left((c_4-c_3)k_-^2 + (\rho_3-2\rho_1) v_R^2 \right) & 0 & 0 \\
		(c_1+c_3) k_1 v_R & (c_1+c_4) k_2 v_R & 0 & 2\rho_1 v_R^2 & \gamma_1 v_R v_S \\
		\beta_1 k_1 v_S -\dfrac{k_1 k_2^2 \Delta'}{v_S} & \beta_1 k_2 v_S -\dfrac{k_1^2 k_2 \Delta'}{v_S} & 0 & \gamma_1 v_R v_S & 2 \alpha_1 v_S^2 + \dfrac{k_1^2 k_2^2 \Delta'}{v_S^2}
		\end{pmatrix}
		\label{new-cp-even}
		\end{eqnarray}
		\normalsize
		where, $k_\pm^2 = k_1^2 \pm k_2^2$ and $\Delta'=\dfrac{(4\lambda_3 k_-^2 + (c_4-c_3)v_R^2)}{2 k_-^2}$.
		
		While, the square of CP-odd mass matrix (${M_i^0}^{2}$)  in $\left\lbrace \xi_1^0,\xi_2^0,\xi_L^0,\xi_R^0,\xi_S^0\right\rbrace$ basis is,
		\begin{eqnarray}
		\begin{pmatrix}
		k_2^2 \Delta'  & k_1 k_2 \Delta' & 0 & 0 & \dfrac{k_1 k_2^2 \Delta'}{v_S}\\
		k_1 k_2 \Delta' & k_1^2 \Delta'  & 0 & 0 & \dfrac{k_1^2 k_2 \Delta'}{v_S} \\
		0 & 0 & \dfrac{1}{2} [(c_4-c_3) (k_1^2-k_2^2) + (\rho_3-2 \rho_1) v_R^2] & 0 & 0 \\
		0 & 0 & 0 & 0 & 0\\
		\dfrac{k_1 k_2^2 \Delta'}{v_S} & \dfrac{k_1^2 k_2 \Delta'}{v_S} & 0 & 0 & \dfrac{k_1^2 k_2^2 \Delta'}{v_S^2} 
		\end{pmatrix}
		\label{new-cp-odd}
		\end{eqnarray}

	Once we diagonalise the above matrix, the three zero eigenvalues of CP-odd mass matrix corresponds to three Goldstone bosons responsible for giving masses to heavy 
	neutral gauge bosons.
		
		Square of the charged scalar mass matrix (${M^{\pm}}^{2}$), in the basis $\left\lbrace h_1^+,h_2^+,h_L^+,h_R^+\right\rbrace$  is the following,
		\begin{eqnarray}
		\begin{pmatrix}
		\dfrac{(c_4-c_3) k_1^2 v_R^2}{2 k_-^2} & \dfrac{(c_4-c_3) k_1 k_2 v_R^2}{2 k_-^2} & 0 & \dfrac{1}{2} (c_3-c_4) k_1 v_R \\
		\dfrac{(c_4-c_3) k_1 k_2 v_R^2}{2 k_-^2} & \dfrac{(c_4-c_3) k_2^2 v_R^2}{2 k_-^2} & 0 & \dfrac{1}{2} (c_3-c_4) k_2 v_R & \\
		0 & 0 & \dfrac{1}{2} (\rho_3-2\rho_1) v_R^2 & 0 \\
		\dfrac{1}{2} (c_3-c_4) k_1 v_R & \dfrac{1}{2} (c_3-c_4) k_2 v_R & 0 & \dfrac{1}{2} (c_4-c_3) k_-^2
		\end{pmatrix}
		\label{new-charged}
		\end{eqnarray}	
	Diagonalisation of the above matrix gives us two massive charged scalars and two massless Goldstones corresponding to a couple of heavy charged gauge bosons.

		We note that, with non-zero $k_1$ and $k_2$, $W$ (and $Z$)-masses get contribution proportional to $\left( k_1^2 + k_2^2 \right)^{1\over 2}$ while $W_L - W_R$ mixing is proportional to $\frac{k_1 k_2}{v_R^2}$ (see Eq. \ref{new_chargegauge}). Experimental limit  on the $W_L - W_R$ mixing \cite{W-mixing} forces one to choose {\em any one} of these vevs to be very small\footnote{e.g. If we set the mixing angle of $W_L$ - $W_R$ at its maximum allowed value, $k_2$ will be of the order of $0.27 ~\rm GeV$, assuming $k_1 > k_2$.} compared to other keeping $\left( k_1^2 + k_2^2 \right)^{1\over 2}$ fixed at 246 GeV. One can then safely assume $k_1 ^2 + k_2 ^2 \approx k_1 ^2 - k_2 ^2 \approx k_1^2$.

		Thus in  small $k_2$  limit, we can rewrite the scalar mass matrices as,
		
		\begin{eqnarray}
		\label{cpeven}
		{M_r^0}^{2}= \begin{pmatrix}
		2 \lambda_1 k_1^2 & 0 & 0 & (c_1 + c_3) k_1 v_R & \beta_1 k_1 v_S \\
		0 & \frac{1}{2}[ 4 \lambda_3 k_1^2  + (c_4 - c_3) v_R^2] & 0 & 0 & 0 \\
		0 & 0 & \frac{1}{2}[(c_4 - c_3) k_1^2 + (\rho_3 -2\rho_1) v_R^2] & 0 & 0 \\
		(c_1 + c_3) k_1 v_R & 0 & 0 & 2 \rho_1 v_R^2 & \gamma_1 v_R v_S \\
		\beta_1 k_1 v_S & 0 & 0 & \gamma_1 v_R v_S & 2 \alpha_1 v_S^2
		\end{pmatrix}
		\end{eqnarray}

		\begin{eqnarray}
		\label{cpodd}
		{M_i^0}^{2}= \begin{pmatrix}
		0 & 0 & 0 & 0 & 0 \\
		0 & \frac{1}{2} [ 4 \lambda_3 k_1^2 + (c_4 - c_3) v_R^2] & 0 & 0 & 0 \\
		0 & 0 & \frac{1}{2} [(c_4 - c_3) k_1^2 + (\rho_3 - 2\rho_1) v_R^2] & 0 & 0 \\
		0 & 0 & 0 & 0 & 0\\
		0 & 0 & 0 & 0 & 0
		\end{pmatrix}
		\end{eqnarray}

		\begin{eqnarray}
		\label{charged}
		{M^{\pm}}^{2}= \begin{pmatrix}
		\frac{1}{2}(c_4 - c_3) v_R^2 & 0 & 0 & \frac{1}{2}(c_3 - c_4) k_1 v_R \\
		0 & 0 & 0 & 0 \\
		0 & 0 & \frac{1}{2}(\rho_3 - 2\rho_1) v_R^2 & 0 \\
		\frac{1}{2}(c_3 - c_4) k_1 v_R & 0 & 0 & \frac{1}{2}(c_4 - c_3) k_1^2
		\end{pmatrix}
		\end{eqnarray}
		
		It is also evident that inspite of setting $k_2=0$ and as a consequence $\mu_{BS}=0$, the elements of scalar mass matrices, mass eigenvalues and mixing matrices practically remains the same as before. It is easy to verify that in the limit $v_R, v_S >> k_1 >> k_2$ (the first inequality arises from the experimental lower limits on heavy gauge boson masses, discussed in a following section), mass matrix defined in Eq. \ref{new-cp-even}  will practically produce the same eigenvalues and mixing among the scalars as has been resulted from  Eq. \ref{cpeven}. Similarly Eqs. \ref{new-cp-odd} and \ref{new-charged} will generate same masses and mixings as Eqs. \ref{cpodd} and \ref{charged} will do respectively.
		
		
		As mentioned above, the masses and mixings among the scalars in $k_2 \neq 0, ~\mu_{BS} \neq 0$ case are nearly same as the $k_2 = 0$ case,
		Although, a non-zero $k_2$ would result into some new couplings among the scalars which are not present in the later case. Some new decay channels will open up for the scalars like $h^0_2$ and $H^0 _S$. However, these new decay modes will  not affect the decay patterns of the physical scalars in a significant way as the decay rates will be proportional to $k_2 ^2$. We will not  discuss them any further. All the following analysis will be done in $k_2 =0$ limit which could be viewed as some special but not phenomenologically different 
		from the more general situation with both $k_1$ and $k_2$ set to non-zero values.

	The scalar potential has $10$ real parameters, $\lambda_1$, $\lambda_3$, $\rho_1$, $\rho_3$, $c_1$, $c_3$, $c_4$, $\alpha_1$, $\beta_1$ and $\gamma_1$. In order to find a set of acceptable values of the physical Higgs boson masses and the potential to be stable at least at classical level, the parameters of scalar potential must obey the following conditions.
	
	\begin{equation}
	\lambda_1,~ (\lambda_1 + 2\lambda_3),~ \rho_1,~ \rho_3,~ (c_1 + c_3),~ (c_1 + c_4),~ \alpha_1,~ \beta_1,~ \gamma_1~ >~ 0
	\end{equation}
	
	The condition that the physical charged Higgs mass squares be positive, demands 
	$$ c_4-c_3 > 0 
	{~\rm and} ~\rho_3 - 2 \rho_1 > 0
	$$
	Values of $(c_4-c_3)$ and $(\rho_3 -2\rho_1)$ can be constrained from a model independent experimental limit of charged Higgs boson mass.
	
	From the CP-even scalar mass matrix we notice that it is effectively a $3\times 3$ mass matrix in $\{h_1^0,h_R^0,h_S^0\}$ basis and thus  difficult to diagonalise analytically. However, one linear combination of $h_1^0, h_R^0$ and $h_S^0$ will be definitely like the SM Higgs boson with mass 125 GeV and having similar properties with this.
	\begin{eqnarray}
	\label{Mh33}
	{M_r^0}^{2}_{3\times 3}= \begin{pmatrix}
	2 \lambda_1 k_1^2 & (c_1 + c_3) k_1 v_R & \beta_1 k_1 v_S \\
	(c_1 + c_3) k_1 v_R & 2 \rho_1 v_R^2 & \gamma_1 v_R v_S \\
	\beta_1 k_1 v_S & \gamma_1 v_R v_S & 2 \alpha_1 v_S^2
	\end{pmatrix}
	\end{eqnarray}
	
	We will denote the eigenstates of  mass matrix (Eq. \ref{Mh33}) by $h^0, H_R^0, H_S^0$.  The rest of the two massive CP-even and two massive CP-odd scalars do not mix with others, and we shall use the same notation  to identify the  mass eigenstates as we have used to define gauge eigenstates. For the charged Higgs sector, the two massive eigenstates will be denoted by  $H_1^\pm$ (which is a linear combinations of $h_1^\pm$ and $h_R^\pm$) and $H_L^\pm$.   
	
	The $3 \times 3$  block of neutral CP-even mass matrix (see mass matrix (\ref{Mh33})), can be diagonalised numerically. We must keep in mind that one of the eigenstates must correspond to the SM-like Higgs boson $h^0$. This implies that the mass eigenvalue and the corresponding eigenvector must be 
	consistent with the measured value of SM Higgs boson mass and its signal strengths to different decay channels at the LHC. We have done 
	a scan over the parameters of the mass matrix (Eq. \ref{Mh33}) over a range keeping the values of $k_1$, $v_R$ and $v_S$ fixed. We will see 
	in the next section that the value of $k_1$ is fixed from the $W$-boson mass, while a lower limit on the values of $v_R$ and $v_S$ can be obtained from the consideration of the masses of heavy gauge bosons $W_R$ and $A'$ arising in this model. While scanning over the 
	parameters we have set the values of $v_R$ and $v_S$ at their lower limits of 14.7 TeV and 13 TeV respectively.
	
\begin{figure}[H]
		\centering
		\includegraphics[height=8cm, width=12cm]{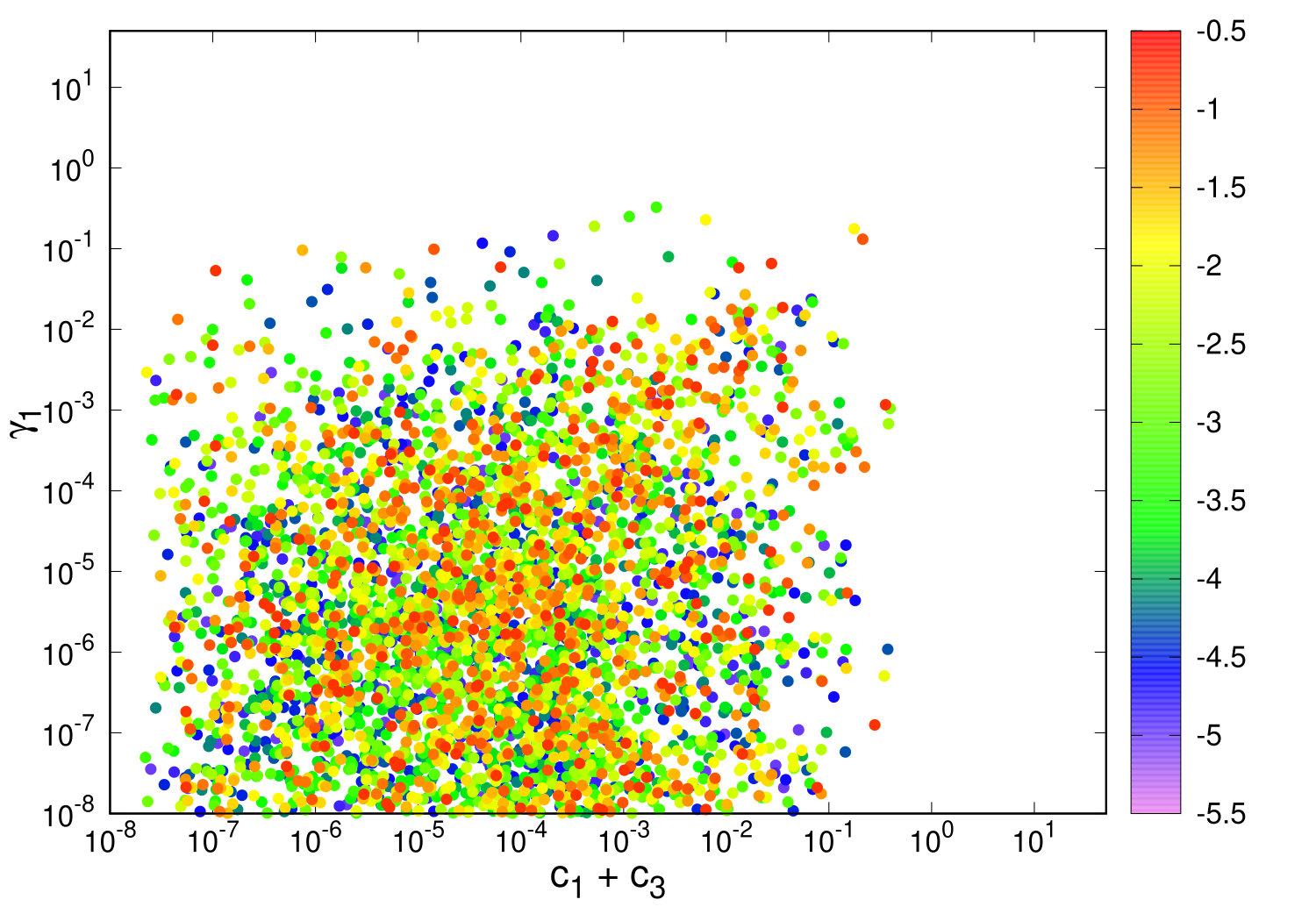}
		\caption{Allowed parameter space for ($c_1+c_3$) and $\gamma_1$ for some fixed values of $\beta_1$. The side bar represents the values of $\beta_1$ in $log_{10}$ scale.}
		\label{beta1limit}
	\end{figure}

	The result of the scan is presented in Fig. \ref{beta1limit}. For the points in the plot, value of one of the eigenstates satisfies the SM Higgs mass condition and Higgs signal strength to $b\bar{b}$ decay mode \cite{W-mixing}. It can be seen from the plot, that relatively larger values of the parameters controlling the off-diagonal terms of  the mass matrix are possible. This in turn implies the eigenstates (particularly the one which can be identified with $h^0$) are linear combinations all three gauge states $\{h_1^0,h_R^0,h_S^0\}$. 
	%
	%
	while performing this scan over a large range of parameter space, we have observed that in most cases it keeps $\lambda_1$ more or less fixed close to the value $\frac{m_{h^0}^2}{2 k_1^2}$. But the values of $\rho_1$ and $\alpha_1$ are completely unconstrained.
	Instead of diagonalising the mass matrix numerically we have restricted ourselves to the values of $c_1 + c_3$ and $\gamma_1$ such that the corresponding off-diagonal terms in the mass matrix can be neglected with respect to the diagonal terms. In this limit, large values of $\beta_1$ forces one to accept large values of $\alpha_1$ so that SM Higgs signal strengths as calculated from the  model is in agreement with experimental observation.
	Furthermore, we keep a tiny value for $\beta_1$ consistent with the above scan result. Under such assumptions about the values of these parameters, mass (squared) eigenvalues can be approximated by the following expressions, 
	
	\begin{eqnarray}
	m_{h^0}^2 & = &  \lambda_1 k_1^2 + \alpha_1 v_S^2 - \sqrt{\Delta}   \nonumber \\
	m_{H_R^0}^2 & \simeq  &2 \rho_1 v_R^2  \nonumber \\
	m_{H_S^0}^2 & = &\lambda_1 k_1^2 + \alpha_1 v_S^2 + \sqrt{\Delta} 
	\label{HSmass} 
	\end{eqnarray}
	with the eigenstate corresponding to the eigenvalue $m_{h^0}^2$ will be identified with the SM-like Higgs boson with mass 125 GeV.
	Here, $\Delta = {(\alpha_1 v_S^2 - \lambda_1 k_1^2)}^2 + \beta_1^2 k_1^2 v_S^2$. Mixing angle, $\theta$, (operative between $h^0$ and $H_S^0$) in small $\beta_1$ limit,  can be written as,
	\begin{equation}
	\tan(2\theta)  =  \dfrac{ \beta_1 k_1 v_S}{\alpha_1  v_S ^2 - \lambda_1 k_1 ^2}
	\label{mixing}
	\end{equation}

	We will be mainly interested  in a study considering in the above mentioned  simplified version of the parameter space where we have only considered that $H^0 _S$ has a tiny mixing (proportional to $\beta_1$) with SM-like Higgs boson $h^0$.  At the end, we will make comment about the possible outcome of a study with non-negligible values of $c_1 + c_3$ and $\gamma_1$.

	\subsection{Gauge sector of $SU(3)_C \otimes SU(2)_L \otimes U(1)_L \otimes SU(2)_R \otimes U(1)_R$ model}
	
	The gauge sector of $32121$ model consists of $16$ gauge bosons namely, the eight gluons ($G^{a'}$, $a' = 1,...,8)$, $SU(2)_{L,R}$ gauge bosons, $W_L^{a}, W_R^{a}, (a=1,2,3)$ and two $U(1)$ gauge bosons $B_L$ and $B_R$. Their interactions are governed by 5 gauge coupling constants $g_3$, $g_{2L}$, $g_{2R}$, $g_{1L}$ and  $g_{1R}$.

	The gauge kinetic Lagrangian can be expressed in terms of field strength tensor in usual way,
	\begin{equation}
	\label{LGK}
	\mathcal{L}_{GK} = -\frac{1}{4} G^{a'\mu \nu} G^{a'}_{\mu \nu} -\frac{1}{4} W_L^{a\mu \nu} W^a_{L\mu \nu} -\frac{1}{4} W_R^{a\mu \nu} W^a_{R\mu \nu} -\frac{1}{4} B_L^{\mu \nu} B_{L\mu \nu} -\frac{1}{4} B_R^{\mu \nu} B_{R\mu \nu} - \frac{\epsilon}{2} B_L^{\mu \nu} B_{R\mu \nu}
	\end{equation}
	
	The last term in Eq. (\ref{LGK}) represents the $U(1)_{L,R}$ kinetic mixing proportional to a dimensionless parameter $\epsilon$.
	A non-zero value of the kinetic mixing coefficient $\epsilon$ would modify the extra heavy neutral gauge boson 
	coupling to a pair of fermions \cite{triparno-U1-mixing}.  However, focus of our present study is not in that direction and we will use $\epsilon =0$ in our following analysis.
	
	The charged gauge bosons mass-matrix (in $W_L - W_R$ basis) follows from the Higgs kinetic Lagrangian (Eq. (\ref{LSK})) in $k_1, k_2 \ne 0$ limit:

		\begin{eqnarray}
		\label{new_chargegauge}
		M_{W^{\pm}}^2 = \frac{1}{4} \begin{pmatrix}
		g_{2L}^2 (k_1^2+k_2^2) & -2 g_{2L} g_{2R} k_1 k_2 \\ -2 g_{2L} g_{2R} k_1 k_2  &  g_{2R}^2 (k_1^2 + k_2^2 + v_R^2)
		\end{pmatrix}
		\end{eqnarray}

	which in small $k_2$ limit appears as,
	\begin{eqnarray}
	\label{chargegauge}
	M_{W^{\pm}}^2 = \frac{1}{4} \begin{pmatrix}
	g_{2L}^2 k_1^2 & 0 \\ 0  &  g_{2R}^2 (k_1^2 + v_R^2)
	\end{pmatrix}
	\end{eqnarray}
	
	Eigenvalues of the  already diagonalised mass matrix  provide  $W_L$ and $W_R$ masses.  The experimentally measured value of $W_L$ mass fixes  $k_1$ at 246 GeV, if we set at $g_{2L} = e/\sin\theta_W$ where, $\theta_W$ is the Weinberg angle. Throughout our article, we shall denote  $W_L$ as the SM $W$ boson with a mass $80.379$ GeV \cite{W-mixing}. 
	Experimentally measured value of $W$ mass  would fix the value of 
	$k_1$ and an experimental lower limit on $W_R$ mass \cite{WRmass} provides a lower bound on $v_R (> 14.7$ TeV).

	One can similarly obtain the mass matrix for  neutral gauge bosons in $ W_{3L}, W_{3R}, B_L, B_R$ basis,
	with $k_2 \ne 0$, $M_{NG}^2$:
		
		\begin{eqnarray}
		\label{new_Zmass}
		\frac{1}{2} \begin{pmatrix}
		\dfrac{1}{2} g_{2L}^2 k_+^2 & -\frac{1}{2} g_{2L} g_{2R} k_+^2 & \frac{1}{6} g_{1L} g_{2L} k_-^2 & -\frac{1}{6} g_{1R} g_{2L} k_-^2 \\
		-\frac{1}{2} g_{2L} g_{2R} k_+^2 & \frac{1}{2} g_{2R}^2 (k_+^2+v_R^2) & -\frac{1}{3} g_{1L} g_{2R} (\frac{1}{2} k_-^2 +v_R^2) & \frac{1}{6} g_{1R} g_{2R} (k_-^2-v_R^2) \\
		\frac{1}{6} g_{1L} g_{2L} k_-^2 & -\frac{1}{3} g_{1L} g_{2R} (\frac{1}{2} k_-^2 +v_R^2) & g_{1L}^2 (\frac{1}{18} k_+^2+\frac{2}{9} v_R^2+\frac{2}{9} v_S^2) & g_{1L} g_{1R} (-\frac{1}{18} k_+^2+\frac{1}{9} v_R^2-\frac{2}{9} v_S^2) \\
		-\frac{1}{6} g_{1R} g_{2L} k_-^2 & \frac{1}{6} g_{1R} g_{2R} (k_-^2-v_R^2) & g_{1L} g_{1R} (-\frac{1}{18} k_+^2+\frac{1}{9} v_R^2-\frac{2}{9} v_S^2) & g_{1R}^2 (\frac{1}{18} k_+^2+ \frac{1}{18} v_R^2 + \frac{2}{9} v_S^2)
		\end{pmatrix}
		\end{eqnarray}
		\normalsize
	which in small $k_2$ scenario practically becomes,
	\begin{eqnarray}
	\label{Zmass}
	M_{NG}^2 = \frac{1}{2} \begin{pmatrix}
	\frac{1}{2} g_{2L}^2 k_1^2 & -\frac{1}{2} g_{2L} g_{2R} k_1^2 & \frac{1}{6} g_{1L} g_{2L} k_1^2 & -\frac{1}{6} g_{1R} g_{2L} k_1^2 \\ 
	-\frac{1}{2} g_{2L} g_{2R} k_1^2  & \frac{1}{2} g_{2R}^2 (k_1^2+v_R^2) & -\frac{1}{3} g_{1L} g_{2R} (\frac{1}{2} k_1^2+v_R^2) & \frac{1}{6} g_{1R} g_{2R} (k_1^2-v_R^2) \\
	\frac{1}{6} g_{1L} g_{2L} k_1^2 & -\frac{1}{3} g_{1L} g_{2R} (\frac{1}{2} k_1^2+v_R^2) & g_{1L}^2 (\frac{1}{18} k_1^2+\frac{2}{9} v_R^2+\frac{2}{9} v_S^2) & g_{1L} g_{1R} (-\frac{1}{18} k_1^2+\frac{1}{9} v_R^2-\frac{2}{9} v_S^2) \\
	-\frac{1}{6} g_{1R} g_{2L} k_1^2 & \frac{1}{6} g_{1R} g_{2R} (k_1^2-v_R^2) & g_{1L} g_{1R} (-\frac{1}{18} k_1^2+\frac{1}{9} v_R^2-\frac{2}{9} v_S^2) & g_{1R}^2 (\frac{1}{18} k_1^2+ \frac{1}{18} v_R^2 + \frac{2}{9} v_S^2) 
	\end{pmatrix}
	\end{eqnarray}
	In practical, the presence of this small $k_2$ will not sensitively affect the masses and mixings in the neutral gauge sector .
		
		Before we make predictions about the masses of the neutral gauge bosons, let us make further assumption about the four gauge coupling constants.
		We will identify the $SU(2)_L$ of $32121$ with the weak isospin group of the Standard Model. It follows automatically that $U(1)_Y$ of the SM will arise due to breaking of $SU(2)_{R} \otimes U(1)_L  \otimes U(1)_R$. Consequently, one can identify $g_Y$ (the $U(1)_Y$ gauge coupling constant with $g_Y = e/\cos\theta_W$) of the SM via the following relation:
		.
		\begin{equation}
		\label{GCValue1}
		\dfrac{1}{g_Y^2} = \dfrac{1}{g_{2R}^2} + \dfrac{1}{g_{1L}^2} + \dfrac{1}{g_{1R}^2}
		\end{equation}
				
		Above relation among the gauge couplings allows us to choose any two of $g_{2R}, g_{1L}$ and $g_{1R}$ independently.
		In order to keep our Lagrangian manifestly LR symmetric, we assume $g_{2L}=g_{2R}$ and $g_{1L}=g_{1R}$. All our analysis presented in the following will be based on this assumption.

To completely determine the gauge boson masses we need to know the values of the gauge coupling constants and three non-zero vacuum expectation values (vevs). The gauge coupling constants have been already fixed from the symmetry breaking condition and demand of manifest Left-Right symmetry. The value or allowed range of values of $v_S$ remains to be known for evaluation of the gauge boson masses from Eqs. (\ref{chargegauge}), (\ref{Zmass}).  It is to be noted that $v_S$ plays a crucial role in breaking $U(1)_L \otimes U(1)_R$. The tree level relation among $m_Z$, $m_W$ and $\cos\theta_W$ has been ensured by identifying the massless eigenstate (of $M^2_{NG}$) with the photon, which has equal coupling to left- and right-chiral fermions.

We have implemented the model Lagrangian in \texttt{SARAH} \cite{SARAH} as well as in \texttt{FeynRules} \cite{feynrules}. In the following 
analysis all the cross-sections will be calculated with help of \texttt{Madgraph5(v2.6.6)} \cite{madgraph} using \texttt{NNPDF23NLO} parton distribution functions \cite{parton-dist} with the factorisation scale set equal to the average mass of the final state particles.
				
Upon diagonalisation, one of the eigenvalues of Eq. (\ref{Zmass}) will give a zero eigenvalue corresponding to the photon. Another eigenvalue comes out to be nearly equal to 91.2 GeV, which we identify with the $Z-$boson. Other two eigenvalues correspond to two heavy neutral gauge bosons which we identify as $Z'$ and $A'$, the last one being a hall mark of an extra $U(1)$ gauge symmetry. 
		
		\begin{figure}
			\centering
			\includegraphics[height=6.5cm, width=10cm]{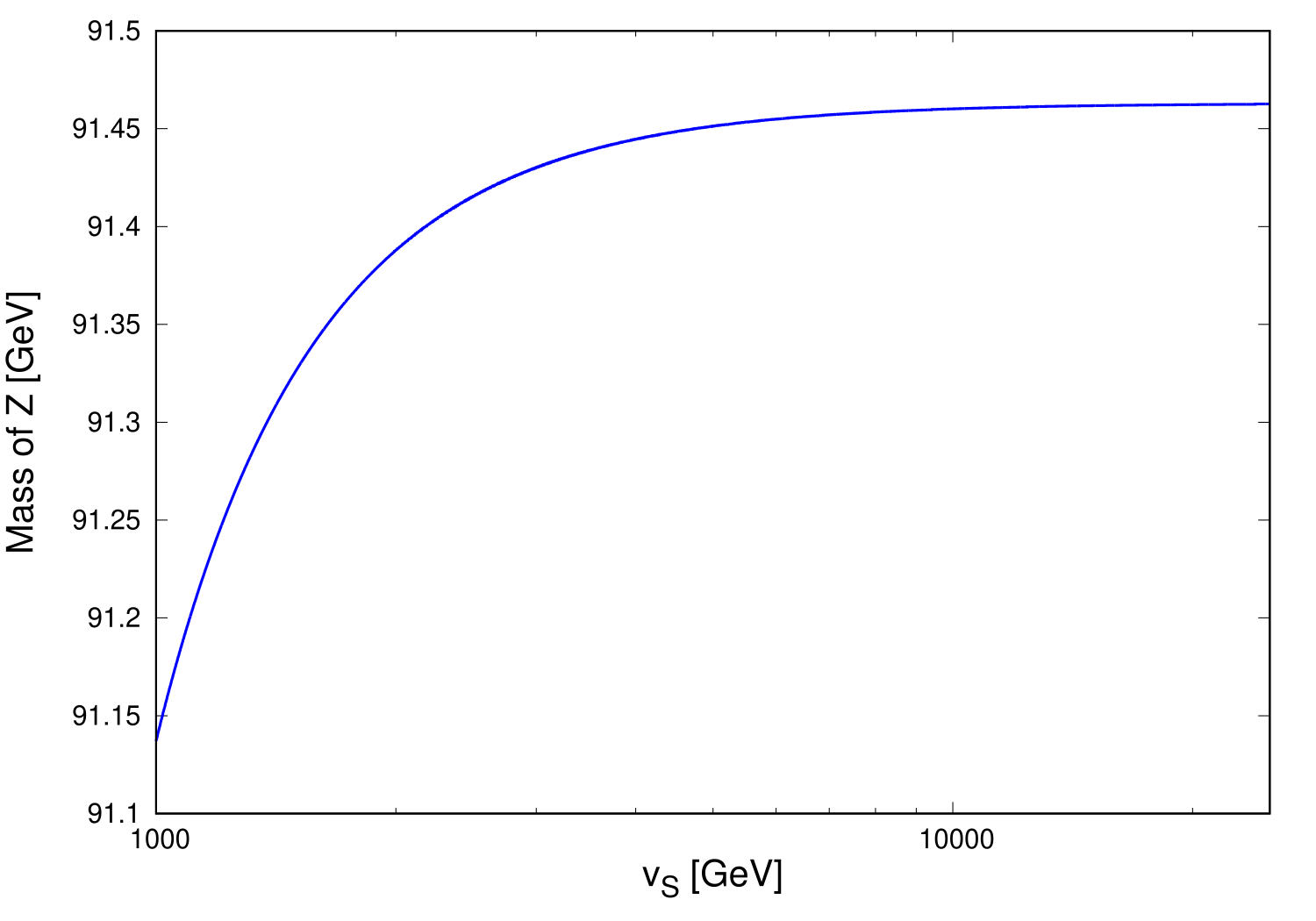}
			\caption{Weak dependence of $Z$ mass on $v_S$}
			\label{MZ_vS}
		\end{figure}
	
\begin{figure}[H]
			\includegraphics[height=6.5cm, width=8.5cm]{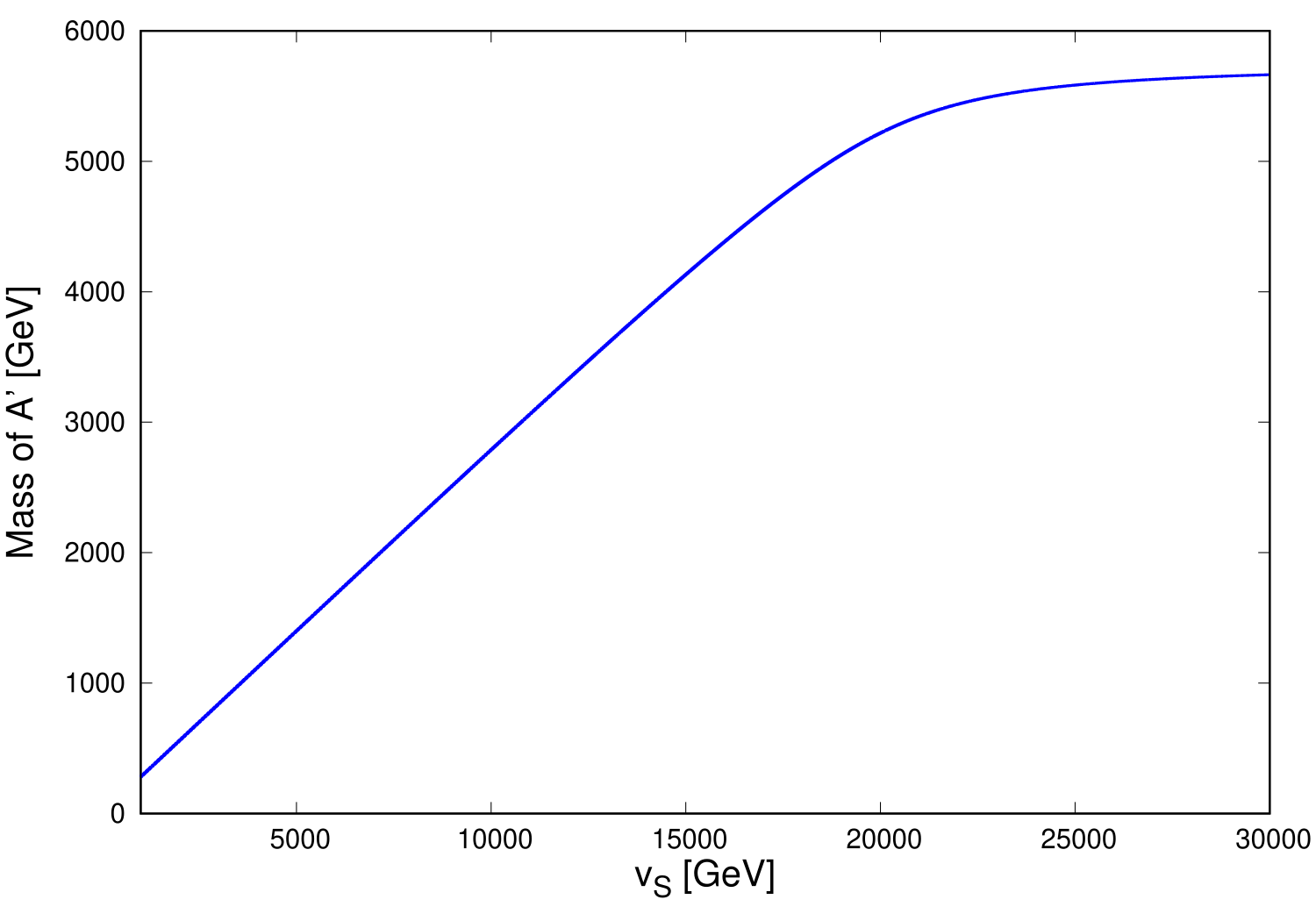}
			\includegraphics[height=6.5cm, width=8.5cm]{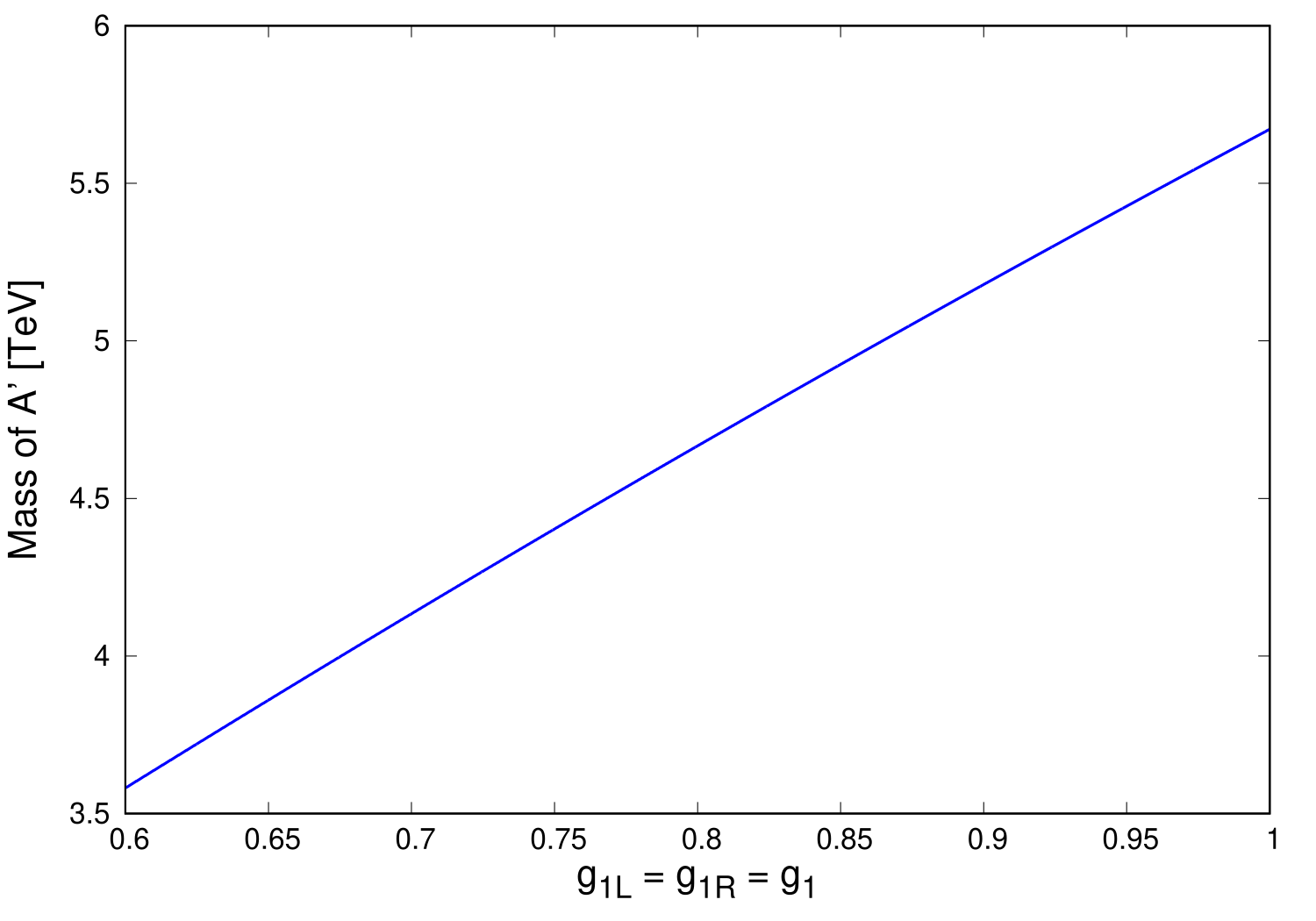}
			\caption{Dependence of $A'$ mass on $v_S$ (left panel) and on $g_{1L}$ and $g_{1R}$, with $g_{1L}=g_{1R}$ for a fixed $v_S$ (right panel).}
			\label{MA1_vS}
		\end{figure}

\begin{figure}[H]
			\centering
			\includegraphics[height=7.5cm, width=11cm]{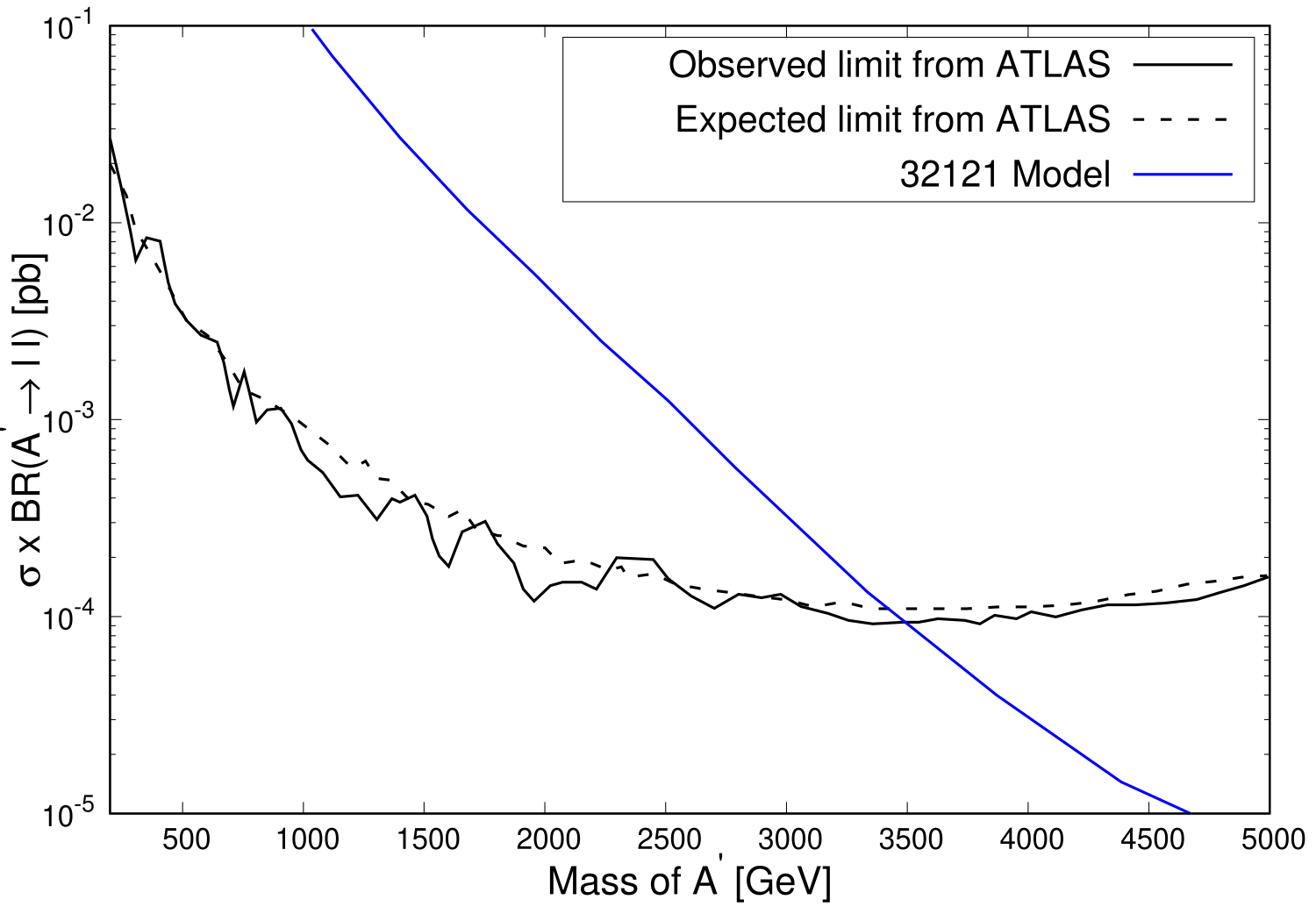}
			\caption{Production cross-section, $\sigma \times BR$ plot for a heavy neutral gauge boson production at the LHC. The black solid (dashed) line represents the observed (expected) 95\% C.L. upper limit on $\sigma \times BR$ from ATLAS (with $\sqrt{s}=13$ TeV, $36.1$ fb$^{-1}$) considering dilepton decay channel of the produced gauge boson and the blue line corresponds to the prediction of $32121$ model.}
			\label{A1bound}
		\end{figure}

	Fig. \ref{MZ_vS} shows the weak dependence of $Z$ mass on $v_S$ whereas, Fig. \ref{MA1_vS} reveals a strong correlation between the mass of $A'$ with $v_S$. Taking a hint from this fact, we would like to find an allowed range of $v_S$ from the LHC data itself. In such an effort, an experimental search of a heavy neutral gauge boson at the LHC and its subsequent decay to a pair of leptons would be helpful. ATLAS collaboration at the LHC \cite{ATLAS-A1} has looked for a pair of high $p_T$ leptons ($e$ and $\mu$) to put an upper limit on the production cross-section times the branching ratio of a heavy neutral gauge boson at 13 TeV. We have translated this upper limit on the $\sigma \times BR$ to the mass of $A'$.
	In our model $A'$ couples to both quarks and leptons with couplings proportional to their $U(1)_{L,R}$ charges.  We present the $\sigma \times BR$ of $A'$ in Fig. \ref{A1bound}. The black solid and dashed lines represent the observed and expected 95\% C.L. upper limit on cross-section times branching ratio by ATLAS respectively. While the blue solid line gives the $\sigma \times BR$ of $A'$ in $32121$ model as a function of $A'$ mass. One can find a lower limit on $A'$ mass equals to 3.5 TeV.
	
\begin{figure}[H]
			\centering
			\includegraphics[height=7cm, width=11cm]{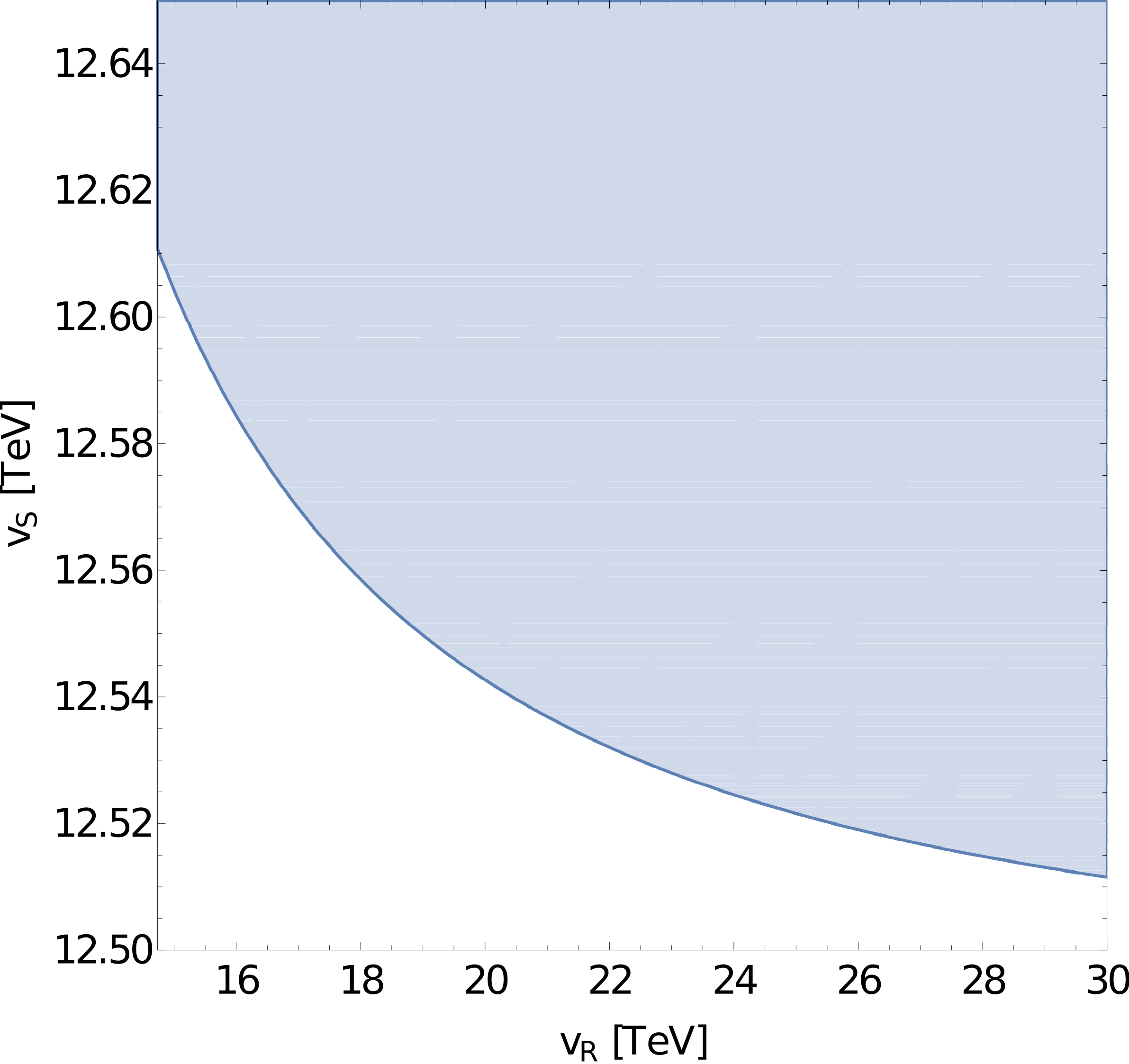}
			\caption{Allowed region in $v_R$ - $v_S$ space, obtained from the limits on $A'$ mass.  $m_{A'}$ has a strong dependence on $v_S$ and have a milder dependence on $v_R$.}
			\label{vrvs-plane}
		\end{figure}

Knowledge of a lower limit on $A'$ mass  enables one to get a lower limit on $v_S$. $m_{A'}$ is a function of the gauge coupling constants and three non-zero vevs necessary for symmetry breaking. Mass of $A'$ has a very weak dependence on $k_1$ and $v_R$ in comparison to $v_S$. Values of the gauge couplings and $k_1$
are fixed. And we set value of $v_R$ at its lower limit while obtaining a lower limit on $v_S$.  We thus arrive a lower limit on $v_S$ which equals to 12.61 TeV.  $m_{A'}$ is a slowly increasing function of $v_R$. So one cannot arrive at an absolute lower limit on $v_S$. The allowed region of $v_R - v_S$ space has been presented in Fig. \ref{vrvs-plane}.
				
		$Z'$ mass on the other hand, is mainly controlled by $v_R$ and it has a weak dependence on $v_S$. With $v_R$ at its lower limit, $Z'$ mass comes out to be 5.9 TeV. For such a massive $Z'$, cross-section times its branching ratio to a pair of leptons is of the order of $10^{-3}$ fb. This rate is well below the upper limit of cross-section times BR for an heavy neutral gauge boson by ATLAS collaboration \cite{ATLAS-A1} and presented in Fig. \ref{A1bound}.

	\subsection{Fermion sector of $SU(3)_C \otimes SU(2)_L \otimes U(1)_L \otimes SU(2)_R \otimes U(1)_R$ model}

The gauge quantum numbers of the fermions have been already listed in Table \ref{table1}. In the following we note the fermions with their chiral components.
	
		\begin{eqnarray}
		L_{L} &=& \begin{pmatrix}
		\nu_{L}  \\ e_{L}
		\end{pmatrix}, \hspace{1.2cm} \hspace{1.3cm} L_{R} = \begin{pmatrix}
		\nu_{R}  \\ e_{R}
		\end{pmatrix} \nonumber \\
		Q_{L} &=& \begin{pmatrix}
		u_{L}  \\ d_{L}
		\end{pmatrix},  \hspace{2.4cm} Q_{R} = \begin{pmatrix}
		u_{R}  \\ d_{R}
		\end{pmatrix}  \nonumber  \\
		Q_{LS} &=& q_{SL}, \hspace{0.5cm} Q_{RS} = q_{SR}, \hspace{0.5cm}  L_S = l_S  \hspace{0.5cm} \mbox{and,}\nonumber  \\
		L_B &=& \begin{pmatrix}
		N_1 & E_1  \\ E_2 & N_2
		\end{pmatrix} \hspace{0.3cm} \mbox{with} \hspace{0.7cm} \bar{L}_B = \begin{pmatrix}
		\bar{N}_1 & \bar{E}_2  \\ \bar{E}_1 & \bar{N}_2
		\end{pmatrix}
		\end{eqnarray}
	
		Here, $L_L,L_R,Q_L,Q_R$ represent SM fermions along with a right-handed neutrino ($\nu_R$). $Q_{LS}$ and $Q_{RS}$ are color triplets and $SU(2)$ gauge singlet exotic quarks having $U(1)_L$ and $U(1)_R$ hyper-charges respectively.  They together form a $4$-component Dirac spinor $q_S$. $N_1$ and $N_2$ are neutral heavy leptons while $E_1$ and $E_2$ are singly charged heavy leptons. They pair-wise form 4-component Dirac spinors, $N$ and $E$ respectively. $l_S$ is a neutral exotic singlet fermion but carrying $U(1)_L$ and $U(1)_R$ gauge quantum numbers.

The fermionic sector of this model consists of several heavy leptons and quarks apart from theirs SM counterparts. We would like to spend few words on them.
The presence of $\nu_R$ facilitates us to write a Dirac or Majorana mass for the neutrinos \cite{Nu_mass1}, \cite{Nu_mass2}\footnote{We have only noted down a possible Dirac mass term for the neutrinos in Eq. (\ref{LRYukawa})}. Heavy charged lepton $E^\pm$ and heavy neutrino $N$ arise from the  $SU(2)$ bi-doublet $L_B$.  These will couple to SM gauge bosons and thus can be produced at the LHC. Similarly, $SU(2)$ singlet quarks $Q_{LS}$ and $Q_{RS}$ form a heavy quark of electric charge $+ {1 \over 3}$ of Dirac type. Finally, there remains a $SU(2)_{L,R}$ singlet lepton of zero electric charge. This could well be a candidate for Dark Matter.  The assignment of $U(1)$ charges for the fermions, from the requirement of anomaly cancellation, is such that the exotic fermions can only couple to the gauge bosons but do not have any mixing with the  SM fermions. This feature will play a crucial role in determining the possible signatures of these fermions at colliders.
		
		Fermions get their masses via their interactions with Higgs fields. The relevant Yukawa Lagrangian is noted below.
		
		\begin{eqnarray}
		\label{LRYukawa}
		\mathcal{L}_{Yukawa} &=& y_{qij} \bar{Q}_{iL} \Phi_B Q_{jR} + \tilde{y}_{qij} \bar{Q}_{iR} \tilde{\Phi}_B Q_{jL} + y_{lij} \bar{L}_{iL} \Phi_B L_{jR} + \tilde{y}_{lij} \bar{L}_{iR} \tilde{\Phi}_B L_{jL} \nonumber \\ 
		&+& y_{sij} \bar{Q}_{iLS} \Phi_S Q_{jRS} + y_{LBij} \; Tr \left[ \bar{L}_{iB} \tilde{L}_{jB} \right] \Phi_S^c + \frac{y_{LSij}}{\Lambda} \bar{L}_{iS} L_{jS}^c \Phi_S \Phi_S \nonumber \\
		&+& y_{BBij}\; Tr\left[ \bar L_{iB} \tilde \Phi_B \right] \L_{jS}^c + H.C.
		\end{eqnarray}
		where, $i,j=1,2,3$ are generation numbers and $y$(s) are Yukawa coupling constants. $\Phi_S^c$ is complex conjugate of $\Phi_S$ and $\tilde{L}_B=\sigma_{2} L_B^* \sigma_{2}$.

In general the Yukawa coupling matrices, $y_{q}$, $y_l$, $y_{LB}, y_s$  are non-diagonal \footnote{In general, in an unifying model like $E_6$, all the Yukawa couplings at the low energy will be generated from a single (and possibly a non-diagonal) Yukawa texture at the
GUT scale \cite{E6fermion}. }. The diagonalisation of the Yukawa matrices in the first line of Eq. \ref{LRYukawa} will give rise to the SM-fermion masses and mixing in the form of $V_{CKM}$ and $V_{PMNS}$. There are no term present in the Yukawa Lagrangian leading to exotic fermion  SM-fermion mixing. Thus while considering the phenomenology of the some of the exotic fermions, we have used their physical masses as the free parameters of the analysis and derive possible bounds on them from LHC itself.
The last term in Eq. \ref{LRYukawa} introduces a mixing between the singlet lepton and the neutral lepton from the bi-doublet. In this work, we shall not be investigating the phenomenological implications of this term.

It is important to note a dimension-4 mass term for the singlet lepton $L_S$ (a Weyl spinor) cannot be written as it transform non-trivially under $U(1)_{L,R}$. 
Using the singlet Higgs field $\Phi_S$, we are able write a dimension-5 operator, which in turn generates  mass for $L_S$. It is well known that any one of the  Higgs bosons  from ${\bf 27}$-plet of $E_6$ cannot give mass to $L_S$. To generate a mass using Higgs mechanism, one must employ a Higgs from a multiplet of $E_6$ other than {\bf 27} \cite{triparno}. So $\Lambda$ may be identified with the vev of such a Higgs boson. Or we can simply assume that $L_S$ has acquired mass from a Higgs belonging to other rep. of $E_6$ and we treat its mass as a free parameter in our analysis.

\begin{figure}
			\centering
			\includegraphics[height=7cm, width=11cm]{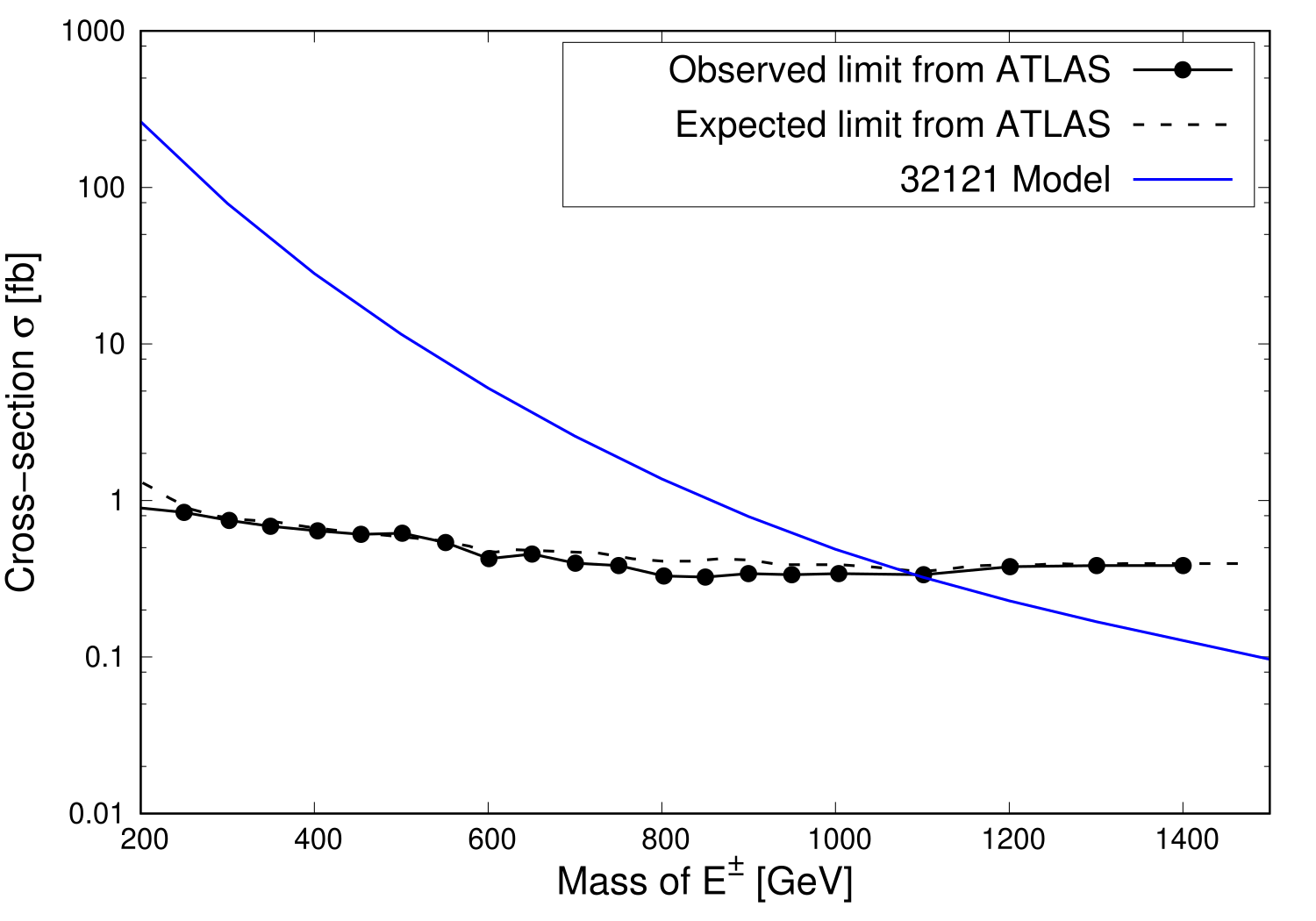}
			\scriptsize
			\caption{Observed (line with dots) and expected (dashed) 95\% C.L. experimental upper limit on the cross-section ($\sigma$) of heavy charged lepton pair-production at the 13 TeV run of the LHC. Also shown in the plot the theoretical prediction from the $32121$ model (blue line).}
			\label{chargedleptonlimit}
		\end{figure}

One can constrain the masses/Yukawa couplings of exotic quarks and leptons from the direct search limits on their masses at the LHC \cite{exotic_quark, exotic_leptons}. 
As for example, ATLAS collaboration has searched for long-lived heavy charged lepton at 13 TeV run with a collected luminosity of 36.1 fb$^{-1}$. We have estimated the pair-production cross-section of $E^+ E^-$ at the 13 TeV and  compared it with the experimental 95\% C.L. upper limit on the same cross-section obtained by ATLAS collaboration. The plots have been presented in Fig. \ref{chargedleptonlimit}. One can see that $E$ mass in 32121 model cannot be less than 1.091 TeV at 95\% C.L.

\section{Phenomenology of  new Higgs bosons  of $32121$ model at the LHC}\label{section3}

Apart from the SM-like Higgs boson, $32121$ model contains a number of neutral and charged scalar states. We will now discuss the possible interactions 
and signatures of such states at the LHC in this section. 

\subsection{Phenomenology of the scalars arising from the bi-doublet  in $32121$ model}

$h_2^0$ ($\xi^0 _2$) is the neutral CP-even (odd) scalar originating from the Higgs bi-doublet, $\Phi_B$. From Eqs. (\ref{cpeven}) and (\ref{cpodd}), we can easily see their masses are equal. They do not decay to a pair of gauge bosons as the vev $k_2$ has been set to zero. For the very same reason, $h^0_2$ or $\xi^0 _2$ does not couple to a pair of any other scalars. 

\begin{figure}[H]
\begin{center}
	\includegraphics[width=10cm, height=4cm]{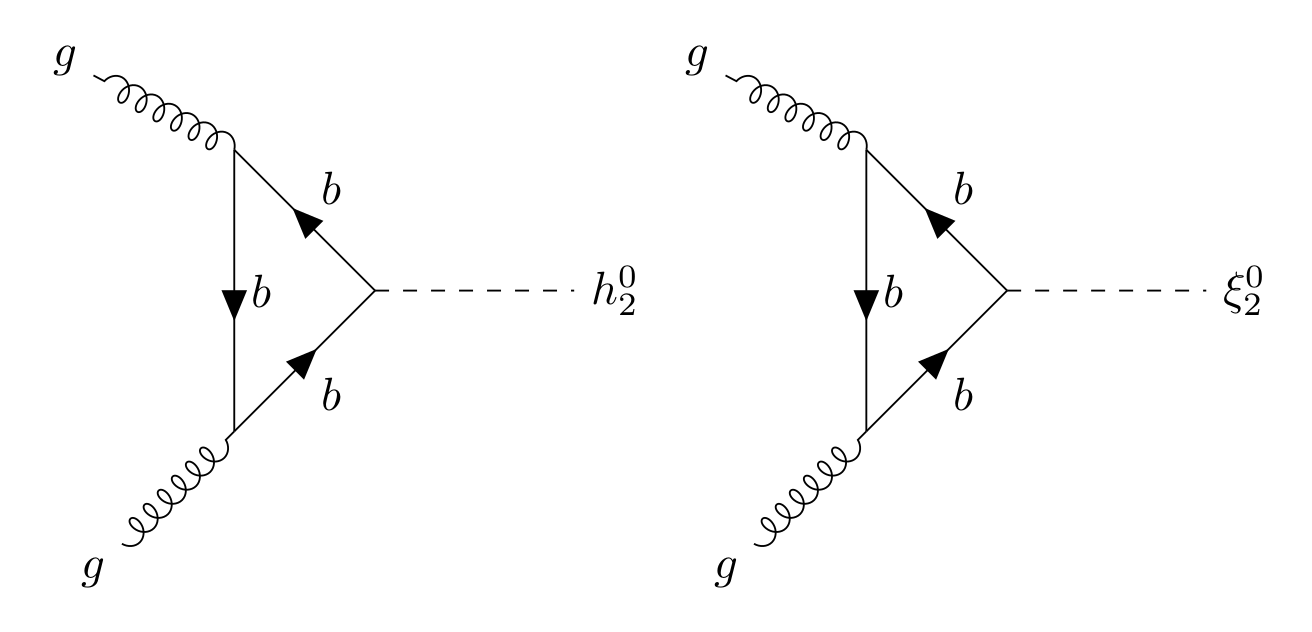}
\end{center}
\caption{Feynman diagrams for $h_2^0$ and $\xi_2^0$ production via gluon gluon fusion through b quark loop}
\label{ggh2}
\end{figure}

\begin{figure}[H]
	\begin{center}
		\includegraphics[width=11cm, height=7cm]{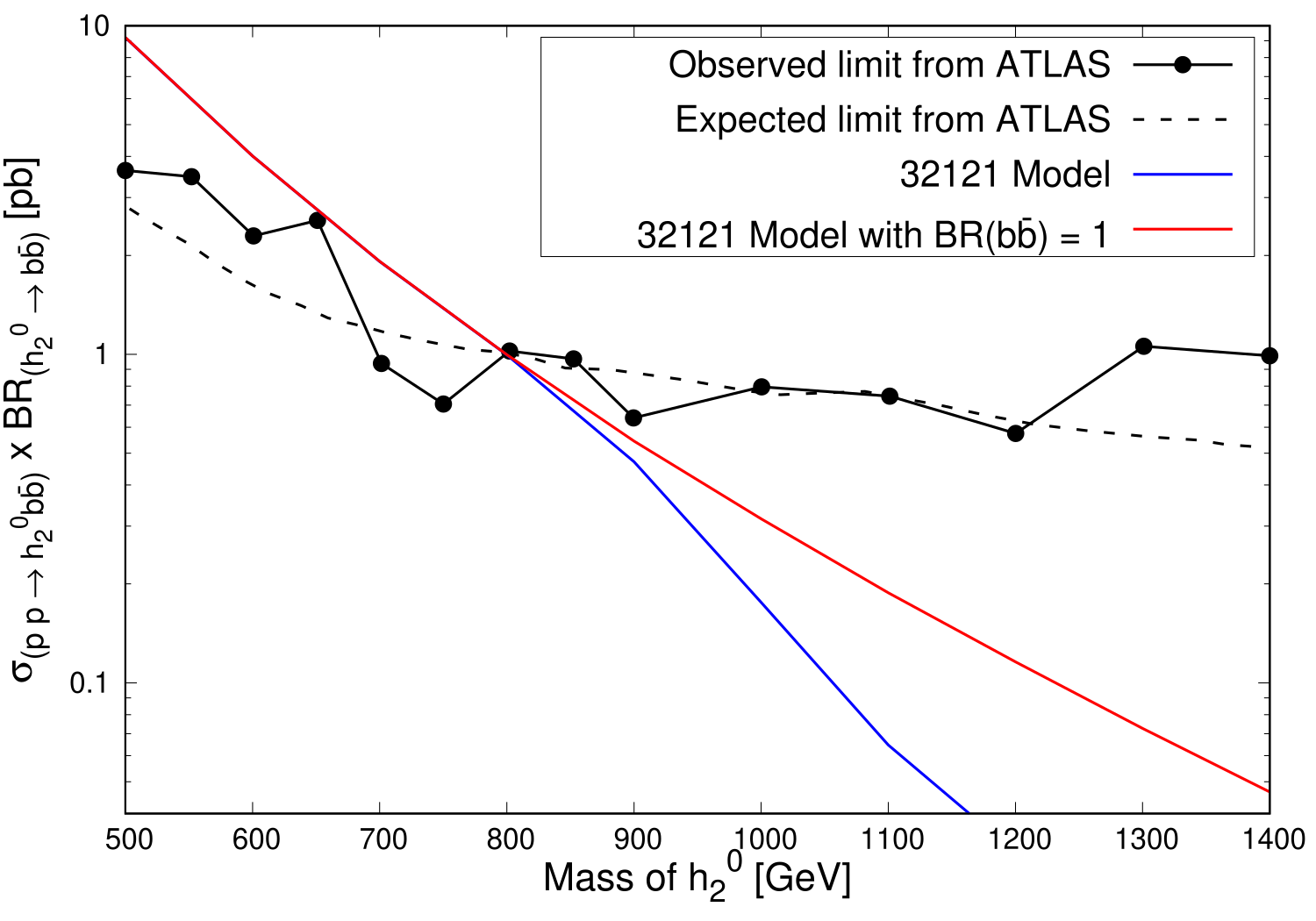}
		\caption{The black solid (dashed) line represents the observed (expected) 95\% C.L. upper limit on the production cross-section ($\sigma$) 
		times branching ratio to $b\bar{b}$, of heavy neutral Higgs boson in association with $b$ quarks as a function of Higgs boson mass at $\sqrt{s} = 13$ TeV with 27.8 fb$^{-1}$ integrated luminosity. The blue line corresponds to $\sigma(p p \rightarrow h_2^0 b \bar{b}) \times BR(h_2^0 \rightarrow b\bar{b})$ whereas the red line represents the same but considering $m_{H_1^\pm} > m{h_2^0}$.}
		\label{h2mass}
	\end{center}
\end{figure}

$h^0_2$ ($\xi^0 _2$) couples to a pair SM fermions via Yukawa coupling (see Eq. (\ref{LRYukawa})). It is interesting to note that the coupling of $h^0_2$ ($\xi^0 _2$) to a pair of top quark is proportional to bottom-Yukawa coupling and vice-versa. ATLAS and CMS have searched for a heavy neutral Higgs boson produced in association with $b$ quarks followed by  its decay to a pair of $b$ quarks at $\sqrt{s} = 13$ TeV \cite{Atlas_h2, Cms_h2}. We consider the production of an $h_2^0$ in association with a pair of $b$ quarks and its decay to a pair of $b$ quarks. The resulting rate of events can be compared with the measured rate by ATLAS Collaboration to set a lower limit on the mass of $h_2^0~(\xi_2^0)$. The calculated (in 32121 model) and (95\% C.L. upper limit on the) measured cross-sections are presented in  the Fig. \ref{h2mass}.  The 95\% C.L. lower limit on $m_{h_2^0/\xi_2^0}$ comes out to be greater than $800$ GeV.  While estimating the $h^0_2$ ($\xi^0_2$) production cross-section in association with a pair of b-quarks, we have incorporated the QCD K-factor ($\sim 1.1$) following the ref. \cite{h2-qcd-k-factor, b_running}. 
However, the lower limit derived in the above, depends on the charged Higgs boson ($H^\pm _1$) mass in the following way. A careful look into the branching ratios of $h^0_2$ reveals that  it dominantly decays to a pair of $b$-quarks, unless a decay to  $H^\pm_1 W^\pm$  is kinematically allowed,. Consequently, mass of $H^\pm_1$ plays a crucial role in determining the rate of $4b$ final state from 
considered above.  A heavier charged Higgs (when $h^0_2 \rightarrow H^\pm _1 W^\mp$ is disallowed)  will push the lower limit on $h^0_2$ mass in upward direction and vice versa. 

\begin{figure}
	\begin{center}
		\includegraphics[width=8.5cm, height=7cm]{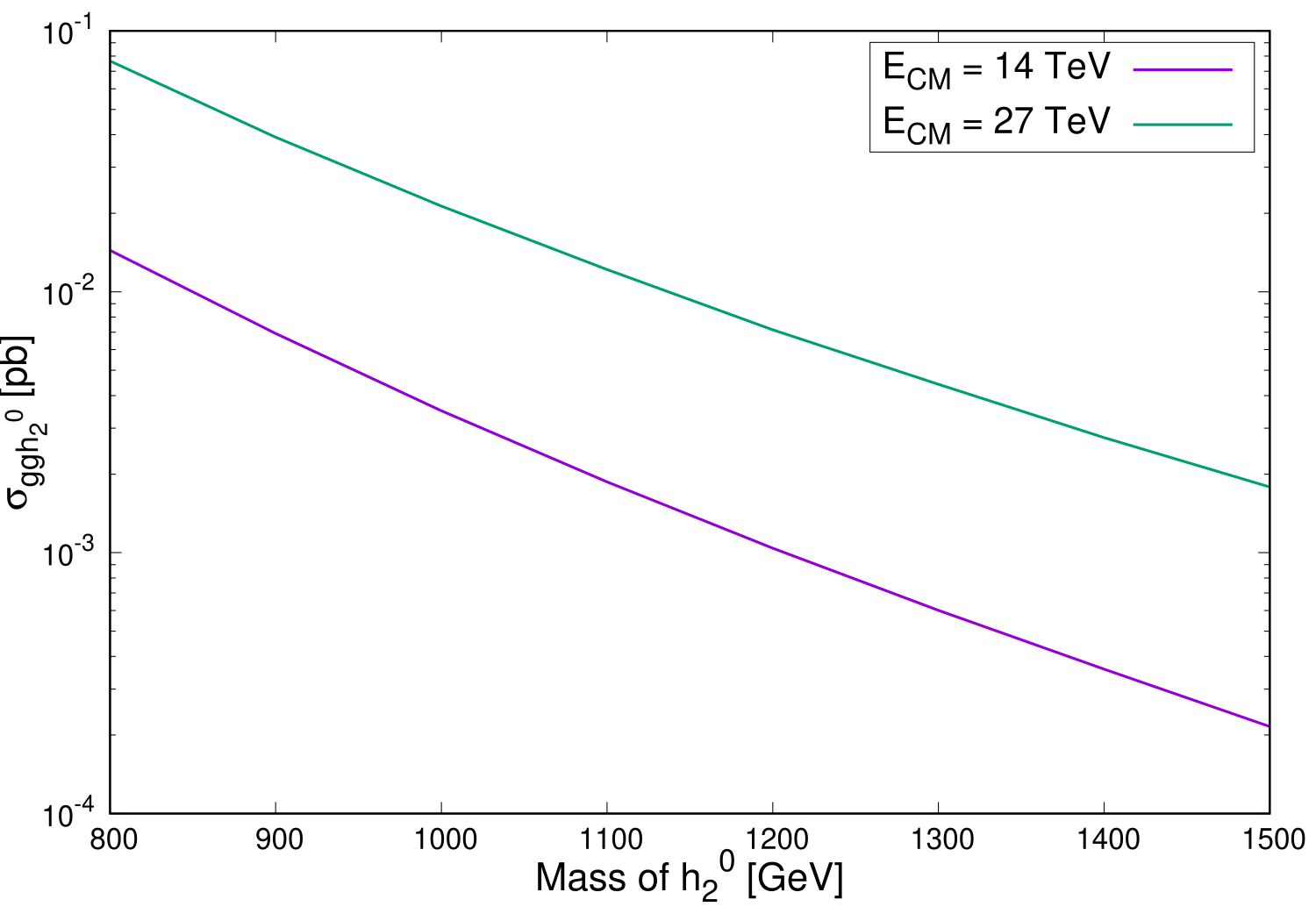}
		\includegraphics[width=8.5cm, height=7cm]{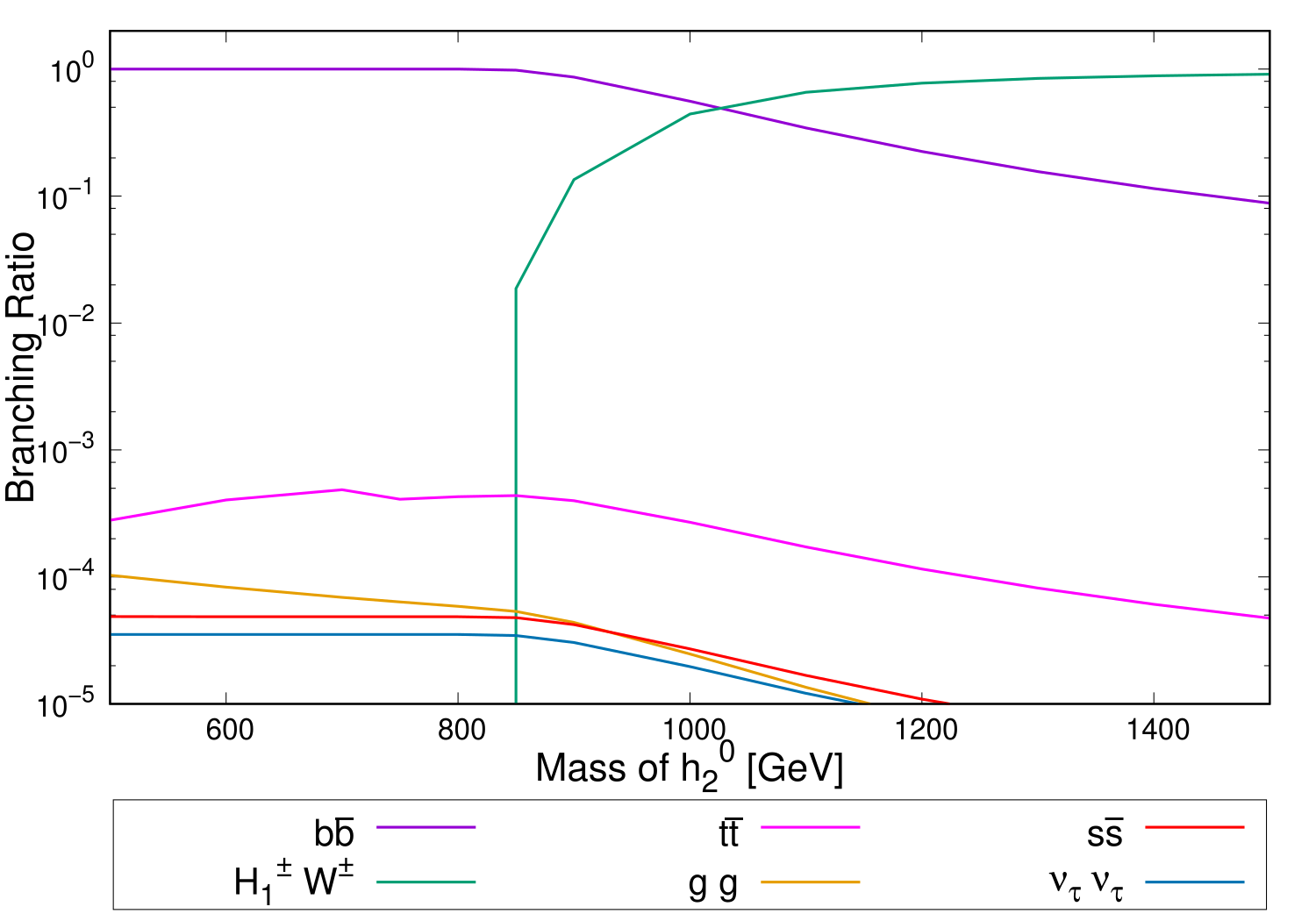}
		\caption{$h^0_2$ production cross section ($\sigma$) via gluon fusion at LHC (left panel) for 14 and 27 TeV proton proton center of mass energy. The right panel shows the branching ratios of $h^0_2$ to different final states. $h^0_2$ and $\xi^0 _2$ have the same masses and coupling strengths}
		\label{h20-prod}
	\end{center}
\end{figure}

In Fig. \ref{h2mass}, we have presented the $\sigma(pp \rightarrow b\bar{b} h^0_2) \rightarrow b\bar{b} (b\bar{b})$ in two cases. The red line represents the rate when $m_{H^\pm_1} > m_{h^0_2}$ and the later decays to a pair of $b\bar{b}$ with 100\% BR. While the blue curve represents the case when $h^0_2$ can also decay to $H^\pm_1$ thus having a reduced decay rate to $b\bar{b}$. A charged Higgs mass of 750 GeV has been assumed while making this plot. 
The  sudden change in the slope of the blue curve due to onset of  $h_2^0 \rightarrow H^\pm _1 W^\mp$ decay mode
around $m_{h^0_2} \simeq 850$ GeV (see Fig. \ref{h20-prod}) is evident.

A dominant production mechanism for such a Higgs boson at the LHC will be via gluon gluon fusion (Fig. \ref{ggh2}). Unlike the SM Higgs boson, in this case, gluon gluon fusion cross-section is dominated by the bottom quark loop. We present the production cross-section (considering NLO QCD correction for this production process, see \cite{qcdcor_ggh}) and decay branching ratios of $h^0 _2$ in Fig. \ref{h20-prod}. 

At the 14 TeV run of the LHC, $h^0_2$ production cross-section varies from 14 fb for $m_{h_2^0} = 800$ GeV to 0.2 fb for 1.5 TeV 
mass of this scalar. Production cross-section at 27 TeV is even higher and it varies from 77 fb at $m_{h_2^0} = 800$ GeV to 1.5 fb
for $m_{h_2^0} = 1500$ GeV. Once produced, $h^0_2$ dominantly decays to a pair of $b$-quarks, unless it decay to $H_1^\pm W^\mp$. 
The later decay mode will only be allowed when $m_{h_2^0}$ is sufficiently higher than $m_{H_1^\pm} + m_{W}$. This choice of mass ordering 
depends on the choice of parameters namely $\lambda_3$ and $c_4 - c_3$. In the plot presented in Fig. \ref{h20-prod}, a certain choice of these parameters have been assumed, so that the $h^0_2 \rightarrow H_1^\pm W^\mp$ has been kinematically allowed, with an additional assumption about the mass of $H_1^\pm$ (750 GeV). 

In the pseudo-scalar sector, $\xi^0 _2$ arises from the Higgs bi-doublet and has a mass equal to the mass of $h_2^0$. It has exactly the similar coupling strengths to the SM fermions as the $h^0_2$. The choice of vanishing $k_2$ forbids its coupling to a pair of gauge bosons or the scalars. Consequently the production and decay mechanism and their rate of $\xi^0_2$ is exactly the same as $h^0_2$. We will not present these numbers separately.

Charged Higgs boson, $H_1^{\pm}$ arises from the Higgs bi-doublet, $\Phi_B$. From the expression (\ref{charged}), one can see, $m_{H_1^{\pm}}^2 = \frac{1}{2} (c_4-c_3) (k_1^2 + v_R^2)$. This massive charged Higgs couples to SM fermions  and decays to a top and a bottom quark with nearly 100\% branching ratio. It also couples to the heavy gauge bosons of $32121$ model but the coupling of $H_1^\pm$ to the SM gauge bosons ($W^\pm, Z$) is proportional to $k_2$, hence identically vanishes. It can be singly produced at LHC with a top and bottom quark or pair-produced via Drell-Yan process or via vector boson fusion process. 

\begin{figure}[H]
	\begin{center}
		\includegraphics[width=11cm, height=6.5cm]{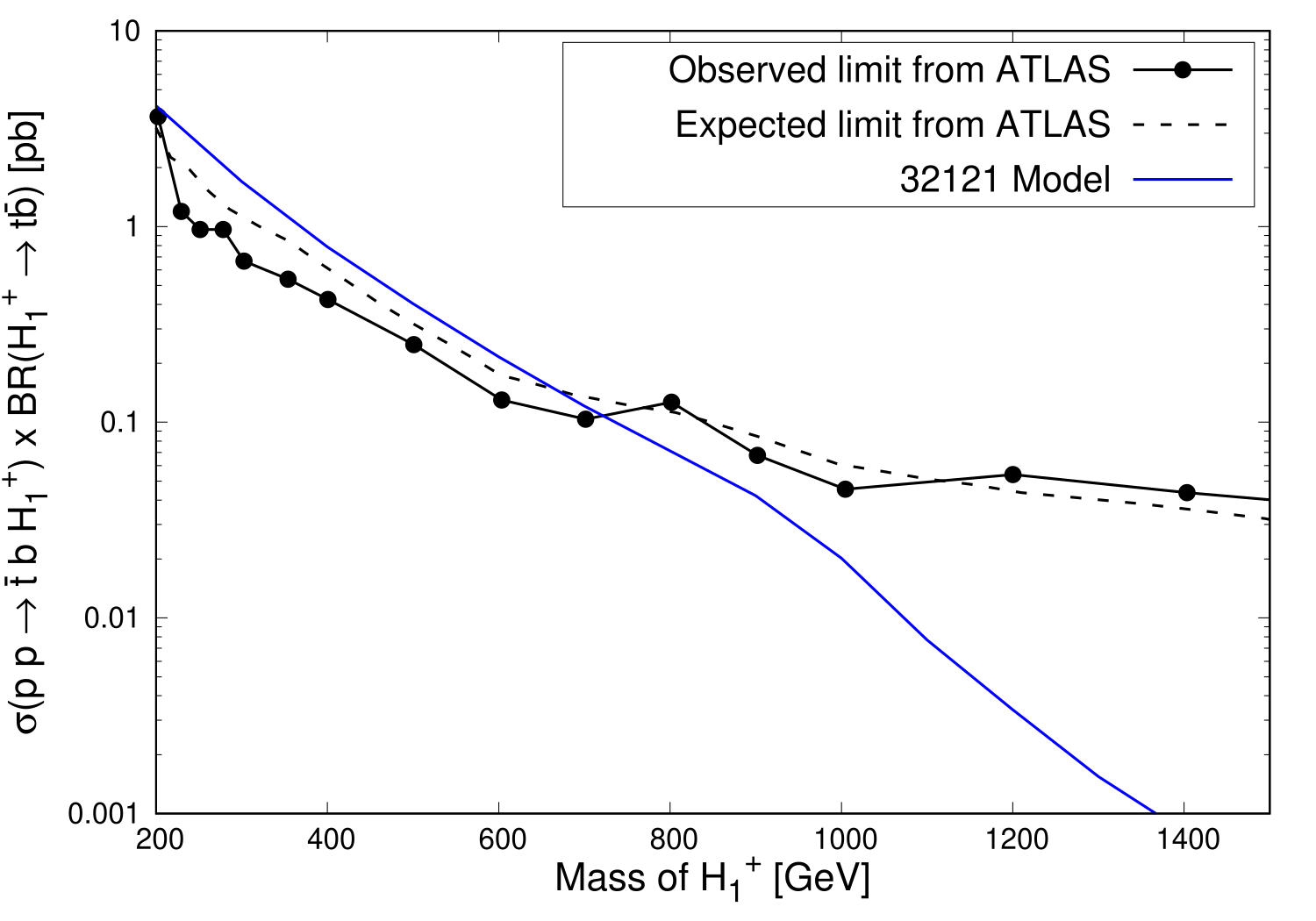}
		\caption{The black solid and dashed line represent observed and expected 95\% C.L. experimental upper limit on the cross-section ($\sigma$) times BR of heavy charged scalar production via $p p \rightarrow \bar{t} b H_1^+ \times BR(H_1^+ \rightarrow t \bar{b})$ at the 13 TeV run of the LHC with 139 fb$^{-1}$ integrated luminosity\cite{AtlasCharged1}. Also shown in the plot the theoretical prediction of $\sigma \times BR$ for $H^+_1$ production in the $32121$ model (blue line).}
		\label{h1p-mass}
	\end{center}
\end{figure}

\begin{figure}[H]
	\begin{center}
		\includegraphics[width=8.5cm, height=8cm]{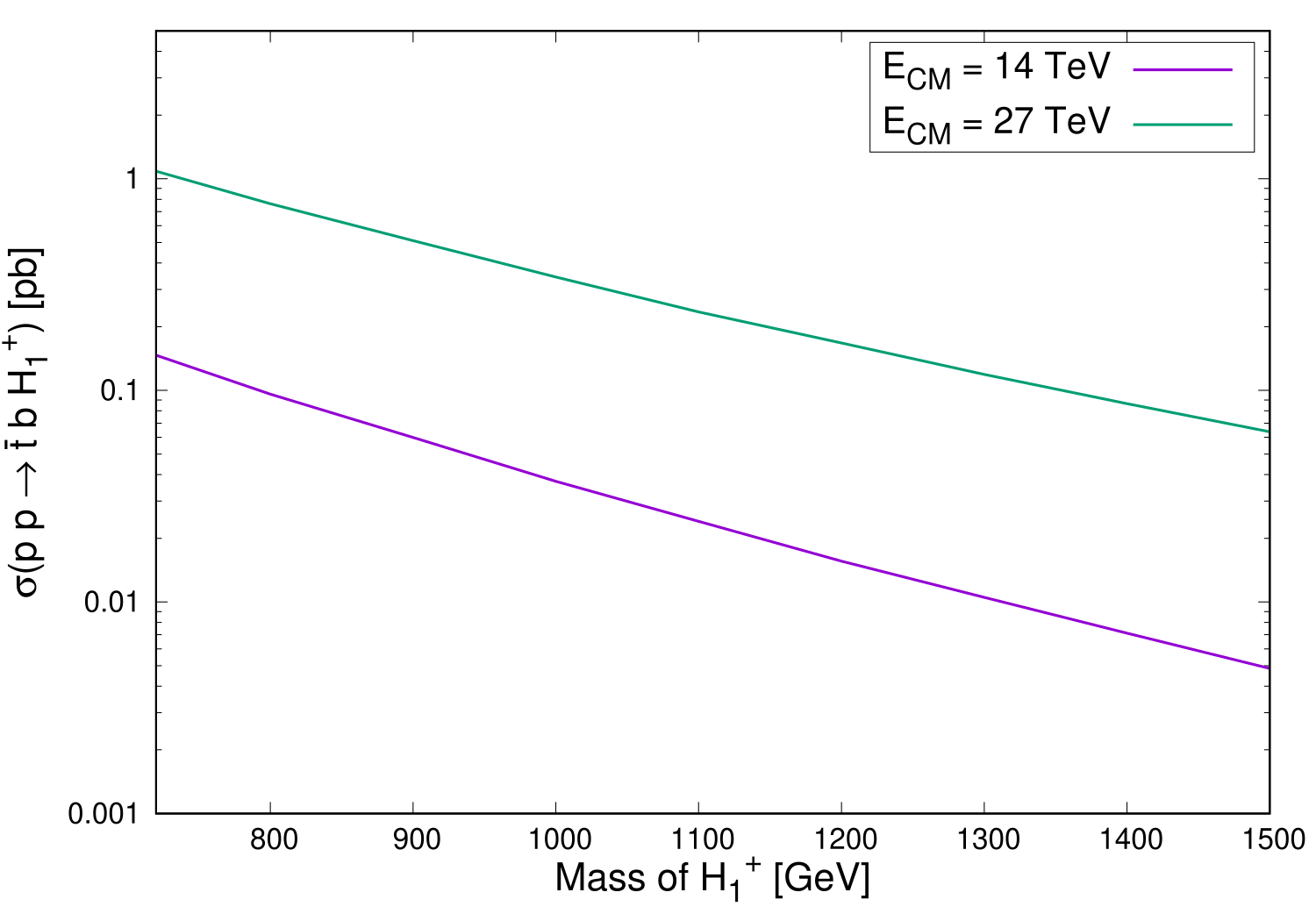}
		\includegraphics[width=8.5cm, height=8cm]{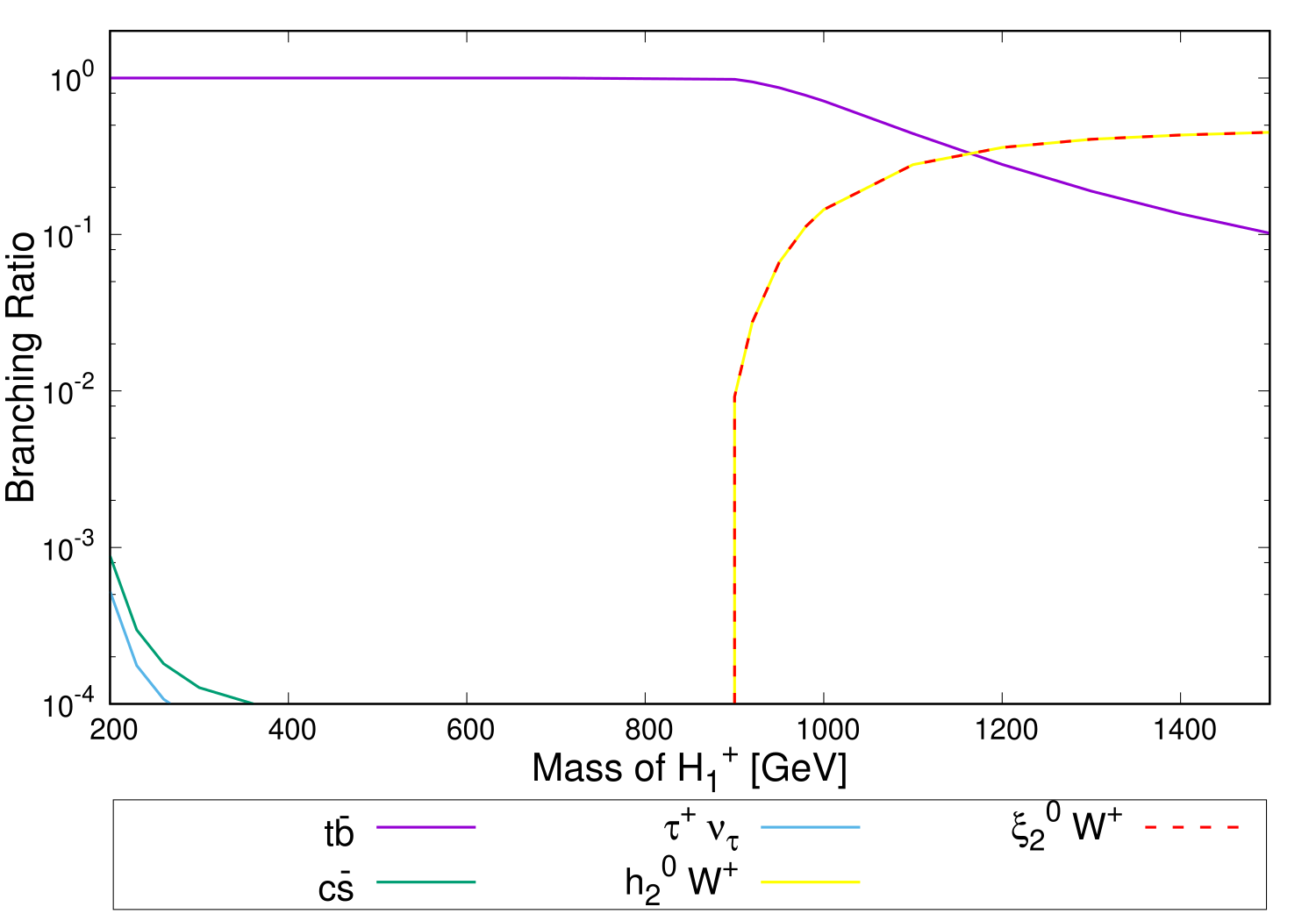}
		\caption{$H_1^\pm$ production cross-section ($\sigma$) via $p p \rightarrow \bar{t} b H_1^+$ process at LHC (left panel) for 14 and 27 TeV proton proton center of mass energy. The right panel shows the branching ratios of $H_1^\pm$ to different final states setting the mass of $h_2^0~(\xi_2^0)$ at its lowest limit ($800$ GeV).}
		\label{h1p-prod}
	\end{center}
\end{figure}

In Fig. \ref{h1p-prod},  we have presented the cross-section of associated production of $H_1^\pm$ with a top and a bottom at the LHC and branching ratio of $H_1^\pm$. In case of $H_1^\pm$ production, the main contribution will be from $g g \rightarrow \bar{t} b H_1^\pm$. ATLAS and CMS collaborations have searched for a heavy charged Higgs boson decaying to a top and bottom at 13 TeV run \cite{AtlasCharged1, AtlasCharged2, CmsCharged}. Using the most recent upper limit on the $\sigma \times BR$ provided by ATLAS, we put a lower limit on the mass of the charged Higgs $m_{H_1^\pm} >$ 720 GeV (see Fig. \ref{h1p-mass}). The sudden change in the slope of the blue curve representing $\sigma_(p p \rightarrow \bar{t} b H_1^+) \times BR(H_1^+ \rightarrow t \bar{b})$ is due to  the sudden decrease of $BR(H_1^+ \rightarrow t \bar{b})$ around $m_{H_1^\pm} = 900$ GeV (see Fig. \ref{h1p-prod}). We have set the mass of $h_2^0~(\xi_2^0)$ at its lower limit of 800 GeV following the Fig. \ref{h20-prod} corresponding to the scenario when $m_{H_1^\pm} > m_{h_2^0}$ (the red line).

We have presented  cross-section for $H^\pm t b$ production (in this process NLO QCD correction and running mass for b quark can be important, see \cite{qcdcor_charged, b_running}) at centre of mass energies of 14 and 27 TeV. In Fig. \ref{h1p-prod}, the right panel shows the branching ratios of $H_1^\pm$ to different final states. Until kinematically allowed for the decay to $h_2^0~W^+$ and $\xi_2^0 W^+$, $H_1^+$ dominantly decays to $t \bar{b}$ ($BR(H_1^+ \rightarrow t \bar{b}) \sim 0.999$). For large $m_{H_1^\pm}$ the branching ratios to $h_2^0~(\xi_2^0) W^+$ channel become more dominant.

The cross-section for $H^\pm tb$ production at centre of mass energy approximately 14 (27) TeV varies from 0.15 (1) pb for $m_{H_1^\pm} = 720$ GeV to 0.005 (0.06) pb for $m_{H_1^\pm} = 1500$ GeV. After being produced, $H_1^\pm$ will decay further and considering respective decay channels (e.g., $tb$ or $h_2^0~(\xi_2^0)~ W^\pm$) one can expect a good amount of events at HL-LHC. However, one needs to consider further decays of top or $h_2^0~(\xi_2^0)$.

\subsection{Phenomenology of the scalars arising from Left-handed Higgs Doublet}

In this section, our primary concern will be  the  neutral and charged states originating from the left-handed doublet 
scalar. Among the neutral CP-even scalar $h_L^0$, neutral CP-odd scalar $\xi_L^0$ and charged scalars $H_L^\pm$, The first two  have equal masses (see Eqs. (\ref{cpeven}), (\ref{cpodd}), (\ref{charged})) and do not mix with other neutral states.  These three states can be pair-produced at the LHC via quark anti-quark fusion mediated by one of the electroweak gauge bosons. However as we set $v_L$ to be zero, neither of these states decays to a pair of SM particles. 

As already pointed out, we will not vary all the parameters of the mass matrix independently to study the masses of the scalars. We will treat the physical masses as free parameters of our analysis. However some caveats are to be imposed on some combinations of parameters of the mass matrices.
As for example, $\rho_3 - 2 \rho_1$ will always assumed to be a positive quantity which is ascertained from the positivity of charged Higgs boson ($H_L^\pm $ ) mass (squared). 
Now the other charged Higgs boson ($H_1^\pm$ ) mass squared is proportional to $c_4 - c_3$. This in turn forces us to take this combination also to be positive. 
As a consequence, masses of $h_L^0$ and $\xi_L^0$ are always greater than mass of $H_L^+$.  However, $(m_{h^0_L} - m_{H^+ _L})$ can be controlled by choosing a proper magnitude of the combination $(c_4 - c_3) \frac{k_1 ^2}{2}$.
 From the expressions (\ref{cpeven}) and (\ref{charged}), $m_{h_L^0}^2 = (c_4 - c_3) \frac{k_1 ^2}{2} + m_{H_L^\pm}^2 $. We will see in the following that $h^0_L$ will decay to $H_L^+ W^-$ if kinematically allowed. So in order to make $h_L^0$ stable, one needs to set $(c_4 - c_3)\frac{k_1 ^2}{2} < m_W^2 + 2 m_W m_{H_L^\pm} $. However, an unstable $h^0_L$ implies that the  mass of $H_1^\pm$ becomes too heavy in the ballpark of 17 TeV.

\begin{figure}[H]
\begin{center}
	\includegraphics[height=7.5cm, width=11cm]{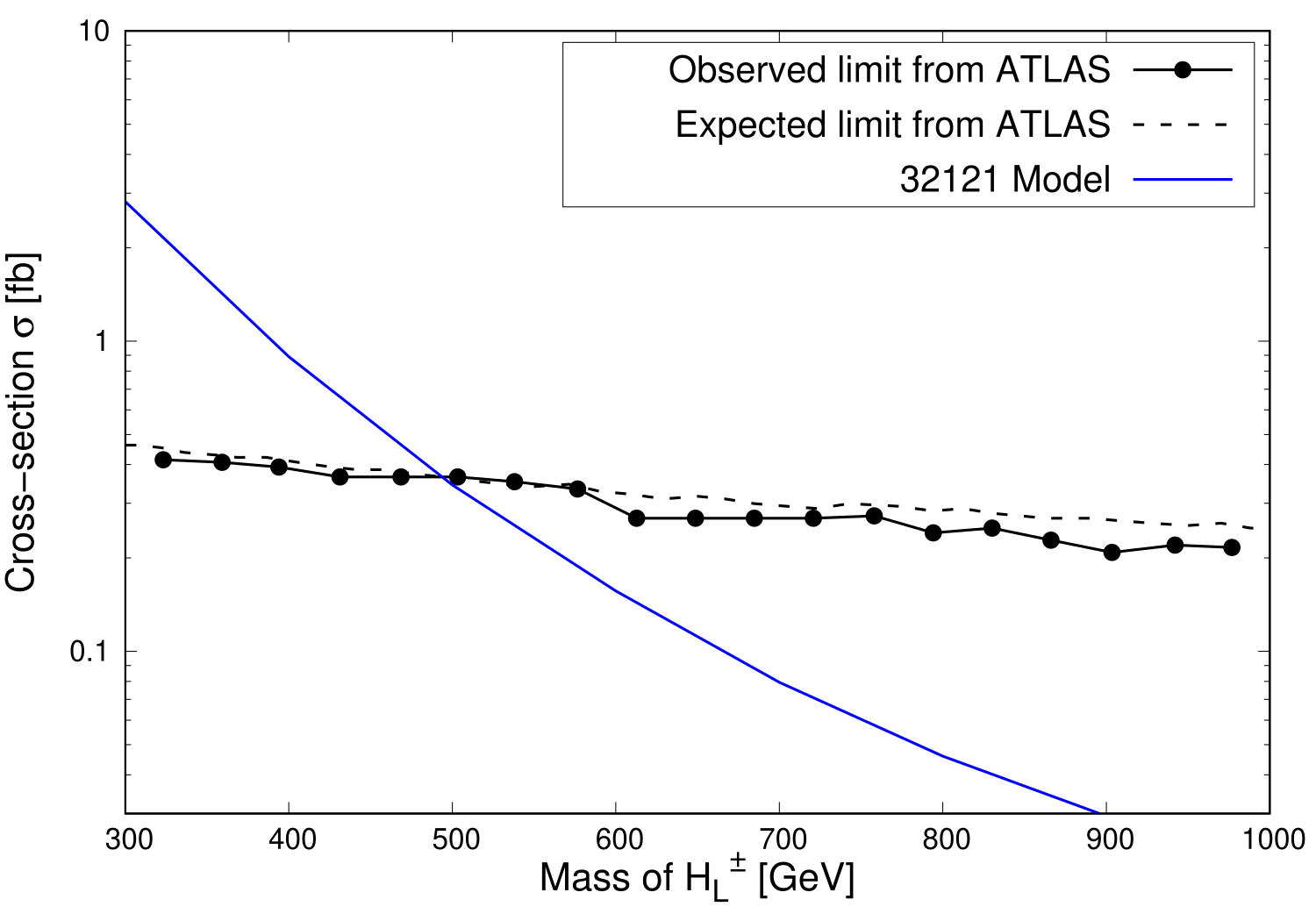}
	\caption{Observed (line with dots) and expected (dashed) 95\% C.L. experimental upper limit on the cross-section ($\sigma$) of heavy stable charged scalar pair-production at the 
		13 TeV run of the LHC. Also shown in the plot the theoretical prediction for $H^+_L H^-_L$ pair-production cross-section in the $32121$ model (blue line).}
	\label{stau}
\end{center}
\end{figure}

 In the following analysis, $h^0_L$ and $\xi^0 _L$ are assumed to be  stable.  Thus they could be potential candidates for DM. $H_L^\pm$ also do not have any decay mode. Once produced at colliders, it passes through the detector without decaying. However being a charged particle, it leaves its signature in the tracker and e. m. calorimeter before leaving the detector.
ATLAS collaboration has searched for long-lived stau $(\tilde \tau$, the super-symmetric partner of $\tau$-lepton) which are very similar to the $H_L^\pm$ \cite{exotic_leptons}. So the upper limit of the cross-section of pair-production of such long-lived $\tilde \tau$s at LHC centre of mass energy of 13 TeV, as quoted by ATLAS collaboration can be used in our case to constrain the $m_{H^\pm _L}$ which is the only free parameter that controls the $H_L^\pm$ pair-production. In Fig. \ref{stau}, we present the variation of $H_L^\pm$ pair-production cross-section (blue solid line) as a function of its mass. Over-layed are the observed and expected upper limits on the pair-production of long-lived stau (black solid and dashed lines). The intersection of these two curves gives us a 95\% C.L. lower limit of 494 GeV, on the left-handed charged Higgs boson ($H_L ^\pm$) mass. 

\begin{figure}
	\begin{center}
		\includegraphics[width=8.5cm, height=7.8cm]{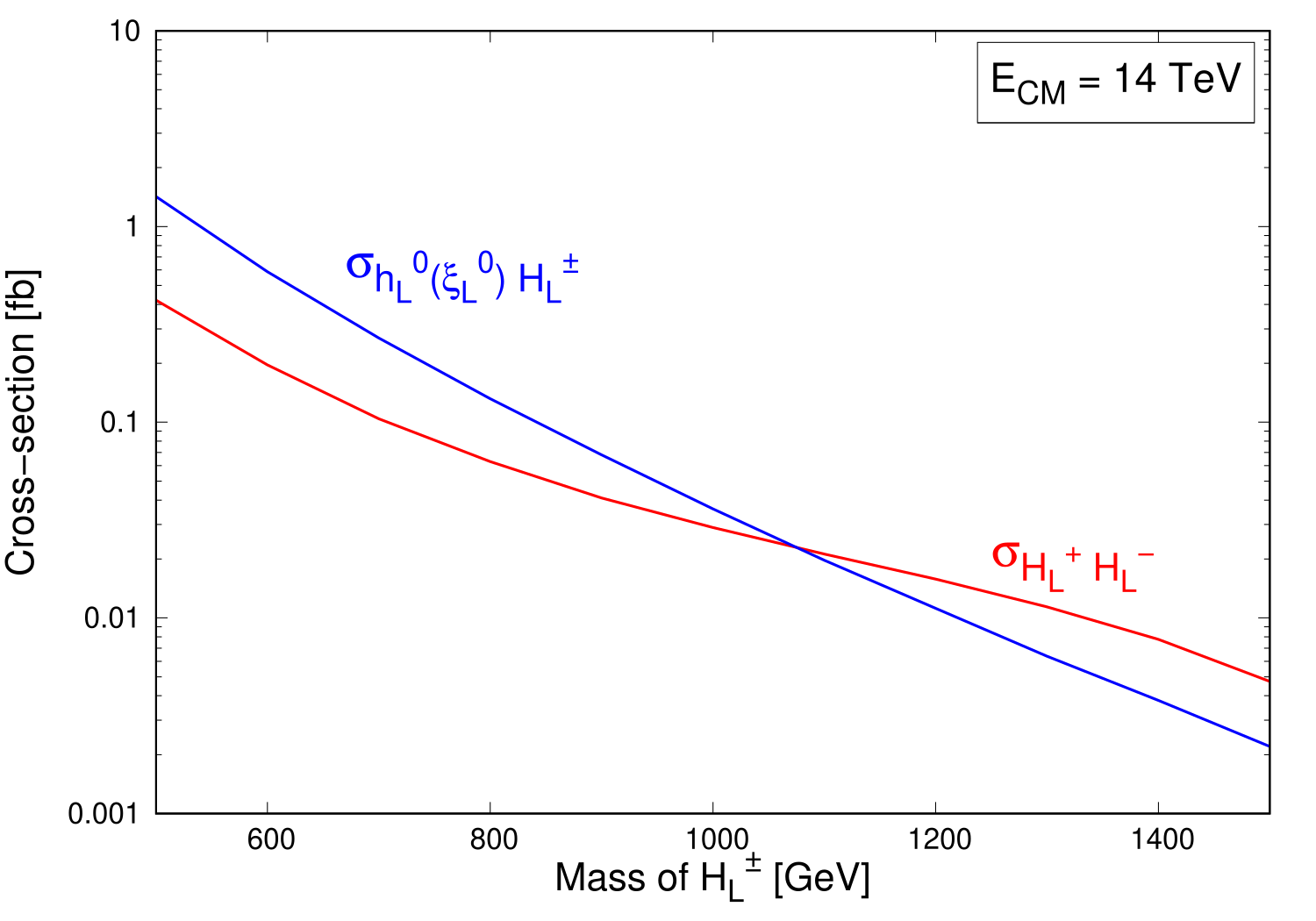}
		\includegraphics[width=8.5cm, height=7.8cm]{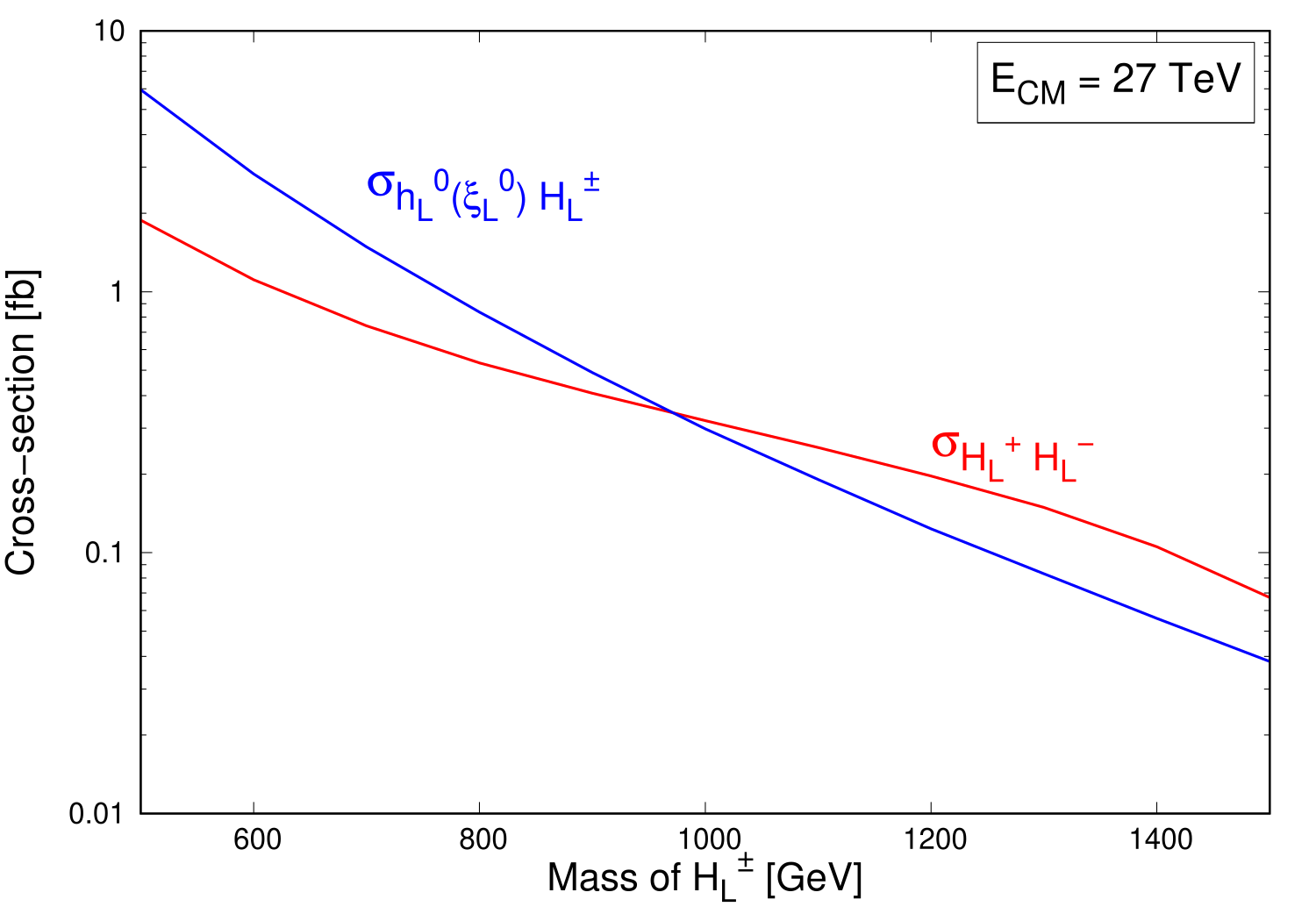}
		\caption{The red solid line corresponds to $H_L^{\pm}$ pair-production cross-section ($\sigma$) at LHC at $\sqrt{s}=14$ TeV (left panel) and at $\sqrt{s}=27$ TeV (right panel) whereas the blue solid line represents the combined production cross-section of one charged ($H_L^{\pm}$) and one neutral (CP-even or CP-odd) scalar at LHC at $\sqrt{s}=14$ TeV (left panel) and at $\sqrt{s}=27$ TeV (right panel)}.
		\label{higgs-prod}
	\end{center}
\end{figure}

Let us now concentrate on the possible production and decay signatures of charged and neutral Higgs bosons arising from the left-handed doublet. As mentioned above, these can be pair-produced at the LHC, via a mechanism similar to Drell-Yan. In Fig. \ref{higgs-prod}, the pair-production cross-sections have been presented with Higgs masses at 14 (27) TeV run of LHC. One can see from Fig. \ref{higgs-prod}, production cross-section for $H_L^\pm$ varies from 0.4 (1.8) fb at 500 GeV to 0.005 (0.06) fb at 1.5 TeV at the center of mass energy 14 (27) TeV. 
$H_L^\pm$ being stable, does not decay any further and  we are left with two ionising tracks of heavy particle in the detector \cite{exotic_leptons}, \cite{HSCP}. At HL-LHC, such a cross-section results into 15 background free events even for a $H_L^\pm$ mass of 1.5 TeV. This particular signature  is unique and cannot arise from the SM.  Thus we hope to explore stable charged Higgs masses upto 1.5 TeV at the 14 TeV HL run of LHC. 

We now turn to the production of a $H_L ^\pm$ in association with either a $h_L ^0$ or $\xi_L ^0$. The production mechanism at the LHC is same as above but 
with a small difference.  The scalar current in the later case is connected to initial state left-handed quark current by a $W^\pm$ propagator. Consequently the 
cross-section for $h_L ^0 H_L^\pm$ is lower than the $H_L^\pm$ pair-production. However, when we combine the $\xi_L ^0 H_L^\pm$ with it, total cross-section 
 of associated production becomes comparable with pair-production rate of charged Higgs bosons. 
 In Fig. \ref{higgs-prod},
associated production cross-section has been presented. One can see that at 14 (27) TeV run of LHC, the cross-section varies from 1.4 (5.9) fb to 0.002 (0.038) fb when the charged Higgs mass varies from 0.5 TeV to 1.5 TeV. 

While discussing the possible signatures of the associated production, we have to be careful about the mass ordering between $H_L^+$ and $h_L ^0~(\xi_L ^0)$. 
When kinematically allowed, $h_L ^0$ will decay (with 100\% branching ratio) to $W^- H_L ^+$. Depending the further decay of the $W$-boson, associated production will result into two charged tracks + 2 jets or 2 charged tracks with a lepton and $E_T\!\!\! \!\!\!/$. On the other hand when $h_L ^0$ is stable, associated production would result into a signal, comprising of a single charged track (from $H^\pm _L$) in association with $E_T\!\!\! \!\!\!/$ \; (arising from $h^0 _L$ and $\xi^0 _L$).

\subsection{Phenomenology of the scalar arising from Right-handed Higgs Doublet}

\begin{figure}
	\begin{center}
		\includegraphics[width=8.5cm, height=8cm]{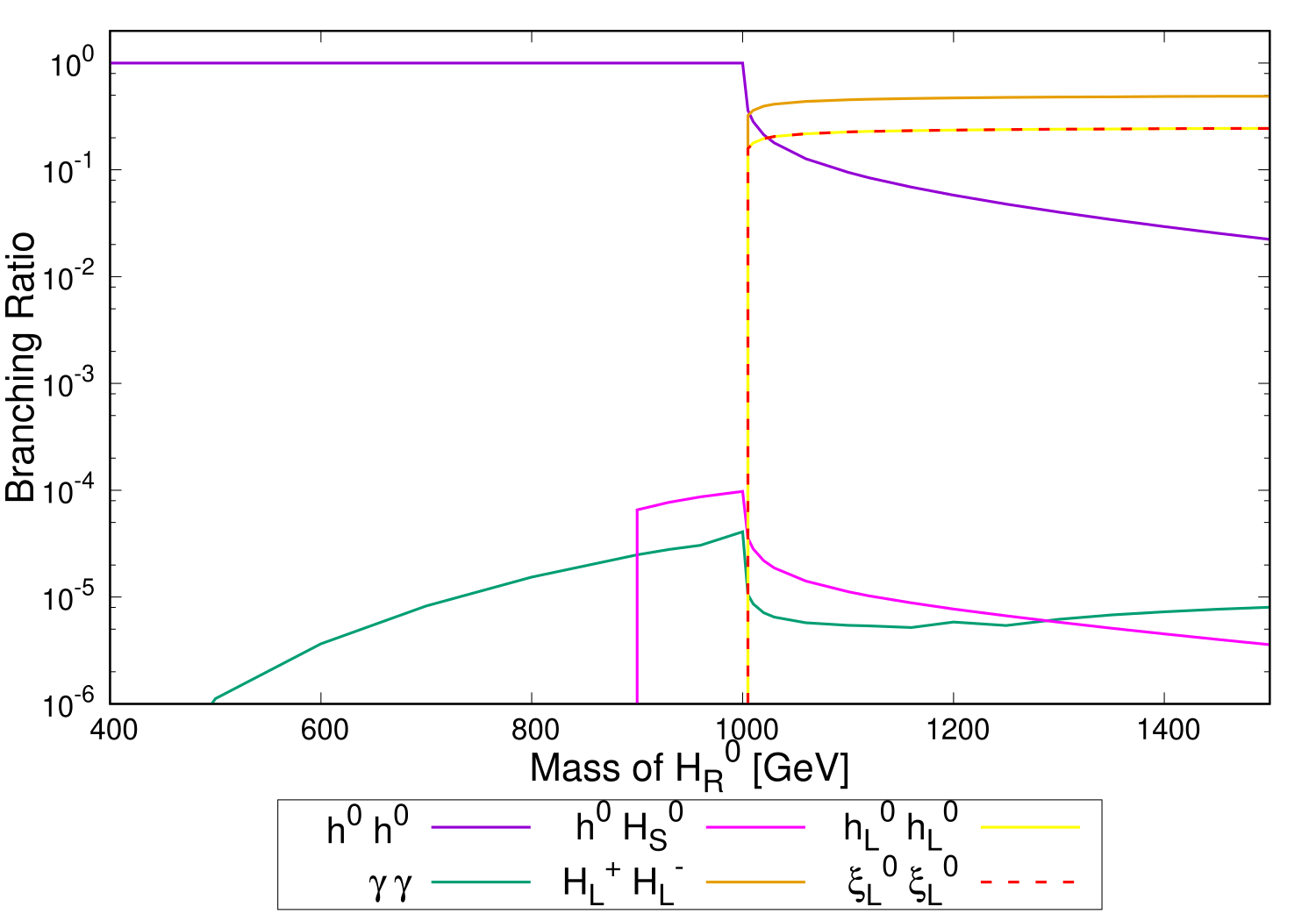}
		\includegraphics[width=8.5cm, height=8cm]{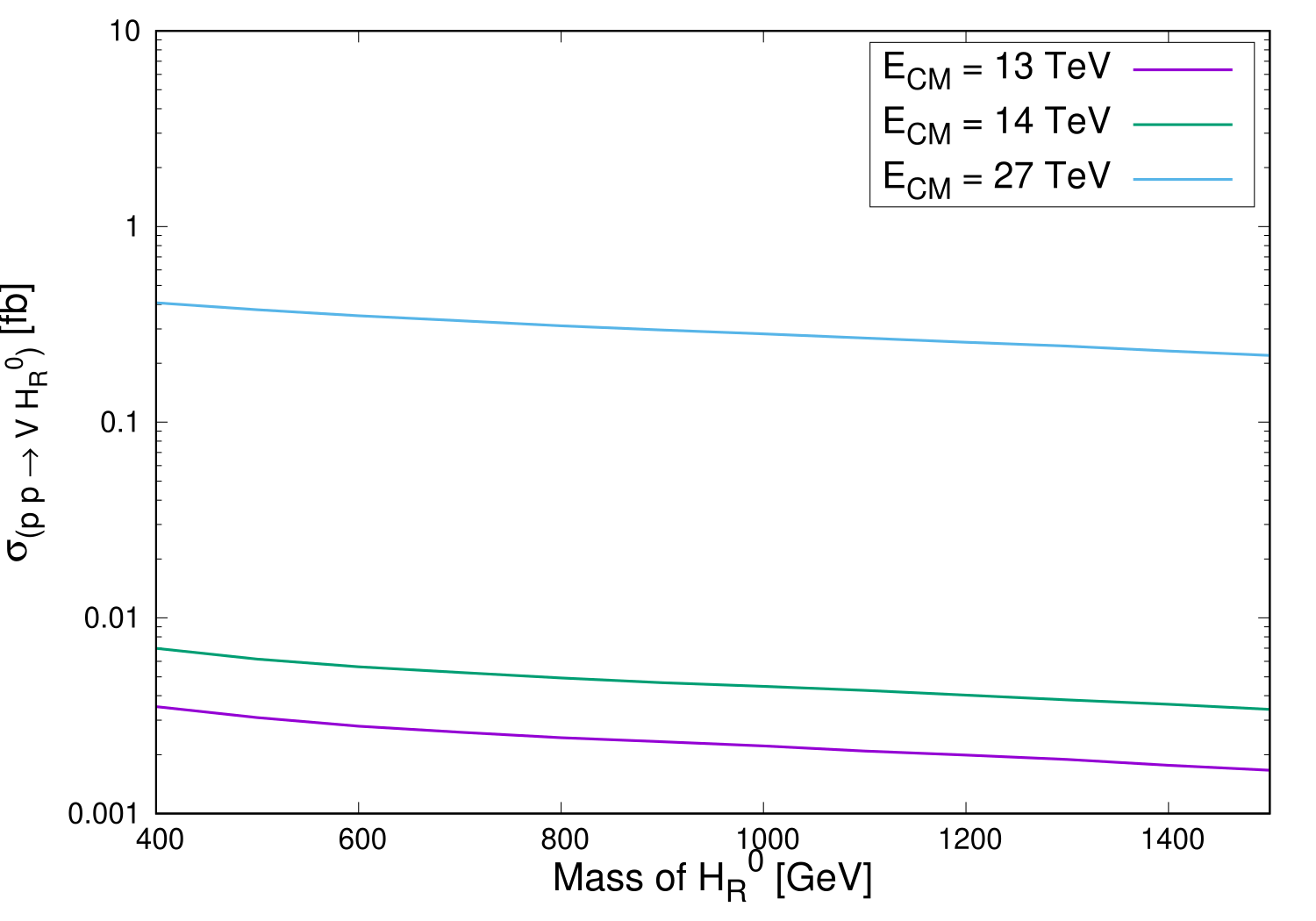}
		\caption{Associated production cross section ($\sigma$) of $H^0_R$ along with a vector boson at LHC (left panel) for 13, 14 and 27 TeV proton proton center of mass energy. The right panel shows the branching ratios of $H^0_R$ to different final states with an assumption of $m_{h_L^0}=m_{\xi_L^0}=500$ GeV and $m_{H_S^0}=700$ GeV.}
		\label{hrbr}
	\end{center}
\end{figure}

Next, in our agenda, is the heavy neutral Higgs boson, $H_R^0$. Due to non-zero $v_R$, it couples to a pair of neutral heavy gauge bosons. But it cannot have any coupling to SM fermions\footnote{It may couple to the SM fermions if we allow a possible mixing between $H^0_R$ with SM Higgs boson}. The plot (Fig. \ref{hrbr}) showing the branching ratios of $H_R^0$  reveals that it dominantly decays to a pair of SM Higgs bosons or to a pair of $H_L^\pm$ or $h_L^0~(\xi_L^0)$ once these decays are kinematically allowed.  Decay to a pair of heavy neutral gauge bosons are  kinematically disallowed. Furthermore, coupling of $H_R^0$ to a pair of $Z$ bosons conspires to be small hence its decay rate to a pair $Z$ bosons is negligible. $H_R^0$ can have an effective coupling at one-loop  ($H^\pm_L$, $W^\pm _R$ and
$H^\pm_1$ running in the triangle loop) to a pair of photons. The  decay branching ratio can be as high as $10^{-5}$ over a wide mass range 
of $H^0 _R$, and is thus not phenomenologically very interesting.
 
\begin{figure}[H]
\begin{center}
	\includegraphics[width=7cm, height=4cm]{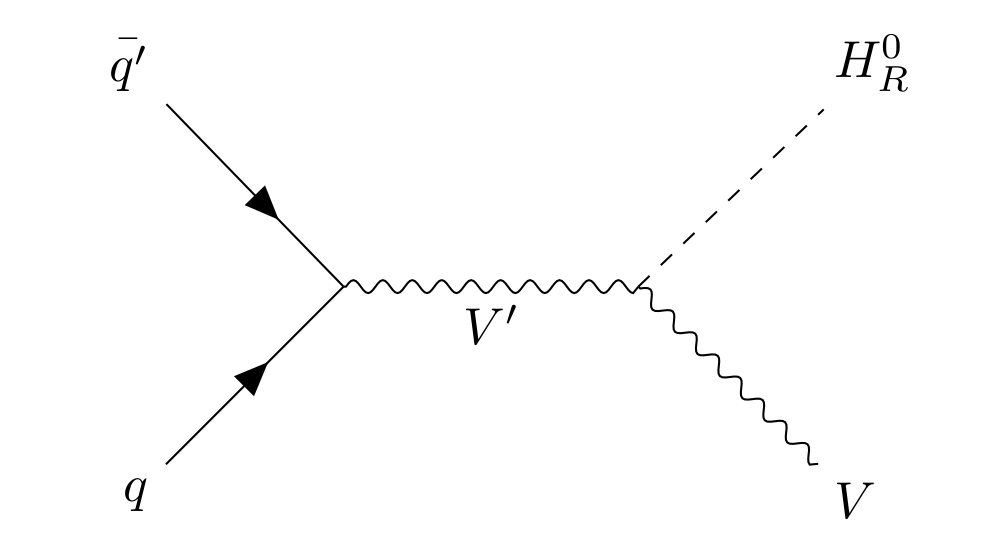}
\end{center}
\caption{Feynman diagram for $H_R^0$ production in association with a vector boson where $V,V'$ represent any vector boson of $32121$ model ($Z$, $Z'$, $A'$ and $W_R$).}
\label{hrdiag}
\end{figure}

The main production mechanism for $H^0 _R$ is  in association with a gauge boson ($Z$, $Z'$, $A'$ and $W_R$) via  annihilation of quark anti-quark pair. It can also be produced in vector boson fusion mechanism. In this article, we will only consider the  production  of $H_R^0$ in association with a vector boson (Fig. \ref{hrdiag}).

In Fig. \ref{hrbr} (right panel) we have presented the combined cross-section of production of a $H^0 _R$ in association with $Z'$, $A'$ and $W_R$, with the heavy gauge boson masses set at their experimental lower limits. Among these three production channels, contribution of $\sigma (H^0 _R A')$ is nearly 70\% of combined  cross-section presented in Fig. \ref{hrbr}. Combined cross-section of $H_R^0$ associated production at the LHC varies from $0.4$ fb to $0.22$ fb for a range of $m_{H_R} : 400$ to $1500$ GeV at a center of mass energy of  $27$ TeV. At $14$ TeV run of the LHC, the cross-section is quite small. It is in the ballpark of $0.005$ fb (Fig. \ref{hrbr}) for a $H^0_R$ of 1 TeV mass. The kinematic suppression due to the presence of a heavy gauge boson in the final state can be one of the reasons  behind the smallness of the total cross-section.

Before we close this sub-section, let us make some brief comments about the possible signature of $H^0_R$ 
at  the LHC. The most promising, signature, in our opinion will arise when $H^0 _R$ can decay to a pair of $H^\pm_L$. As mentioned before, if will produce two charged tracks in the detector with their invariant mass peaking at the mass of $H^0_R$. Along with a pair of charged tracks heavy gauge boson decay will probably give rise to a pair of high mass jets or leptons. As for example, at 27 TeV run of the HL-LHC, 
one expects around 30 two charged tracks two lepton events for a $m_{H^0 _R} = 1$ TeV\footnote{The branching ratio for di-lepton decay ($e,\mu,\tau$) of $A'$ is $\sim 10\%$ and branching ratio of $H_R^0 \rightarrow H_L^+ H_L^-$ is around 50\%. So considering almost $70\%$ contribution of $\sigma(H_R^0 A')$, at $27$ TeV run with $3000$ fb$^{-1}$ integrated luminosity, for the process $\sigma(p p \rightarrow H_R^0 A') \times BR(A'\rightarrow ll) \times BR(H_R^0 \rightarrow H_L^+ H_L^-)$ one can expect around 30 events for 1 TeV $H_R^0$ mass.}. While at the 14 TeV run with high luminosity option, detection of such events seems to be very challenging even for a 500 GeV $H^0 _R$.

\subsection{Phenomenology of the $SU(2)_L \otimes SU(2)_R$ Singlet scalar in $32121$ model}	
		
		Next, in our agenda, is the scalar arising from $SU(2)_L \times SU(2)_R$ singlet $\Phi _S$. $\beta_1$ being small in order to 
		satisfy the SM Higgs bosons signal strength (see Fig. \ref{beta1limit}), it does not have any significant role in the phenomenology of the singlet scalar and 
		 we can treat mass of the singlet scalar itself as the free parameter of our analysis. 
		In the following we intend  to study the decays and dominant production channel of the singlet scalar boson.
		
\begin{figure}[H]
			\centering
			\includegraphics[height=8cm, width=11.5cm]{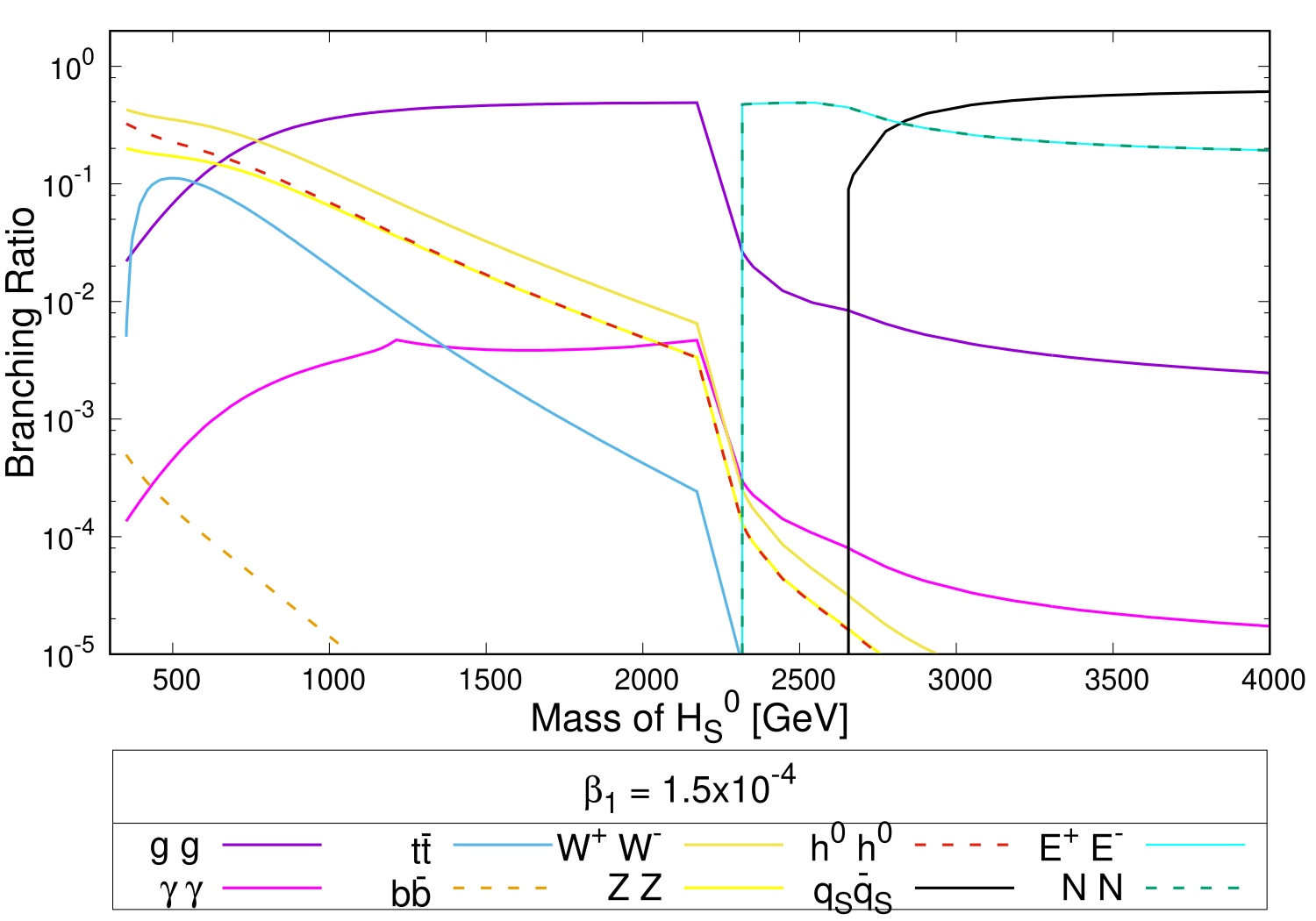}
			\scriptsize
			\caption{Branching ratio of $H_S^0$ to different channels including exotic quarks and exotic leptons for  $\beta_1=1.5\times 10^{-4}$ and $v_S = 13$ TeV and exotic quark mass 1.3 TeV.}
			\label{hsbr}
		\end{figure}

In Fig. \ref{hsbr}, we present branching ratios of $H_S^0$ to its available decay channels. To remind, $WW$, $ZZ$, $t\bar{t}$, $b\bar{b}$  decays of $H_S^0$ take place via the mixing with the SM Higgs boson.  While rest of the decays are driven via the direct couplings of $H_S^0$ to the decaying particles.
		
The dominant contribution to $H_S^0 \rightarrow gg, \gamma \gamma$ decay,  arise from triangle loops of heavy exotic quarks and leptons. Charged Higgs states arising from $\Phi_B$, $\Phi _L$ also  contribute to singlet Higgs decay  to $\gamma \gamma$.  $H_S^0 \rightarrow gg$ is important as production cross-section of $H_S^0$ via gluon fusion is directly proportional to this decay width. However, $v_S$ being large, singlet Higgs Yukawa to exotic leptons/quarks are tiny  (see Eq. (\ref{LRYukawa}) in section 2.3). Consequently, decay width to $gg$ is small.  Similar arguments can be given to understand the smallness of $H_S^0 \rightarrow \gamma \gamma$ decay rate.

The branching ratios to several decay channels are moderately sensitive to $\beta_1$.  With a higher value of $\beta_1  (\sim 10^{-3})$  one can 
satisfy all the constraints from SM Higgs boson signal strengths. However, $\beta_1 > 10^{-3}$ \footnote{$\beta_1 > 0.01$ is excluded
as the singlet component in SM-like Higgs will be too high to satisfy the experimentally measured signal strengths.} will lead to a singlet Higgs boson mass of
700 GeV and above. Furthermore, a higher value of $\beta_1$ leads to a larger mixing between the singlet and the SM-like Higgs boson. 
Thus the singlet Higgs boson decay rates to $tt$, $bb$, $WW$ and $ZZ$ channels will increase slightly. 
The variation of branching ratios over a wide  mass range of $H_S^0$ for a fixed $\beta_1$ have been shown in the Fig. \ref{hsbr}.

\begin{figure}
		\begin{center}
		\includegraphics[width=10cm, height=4.5cm]{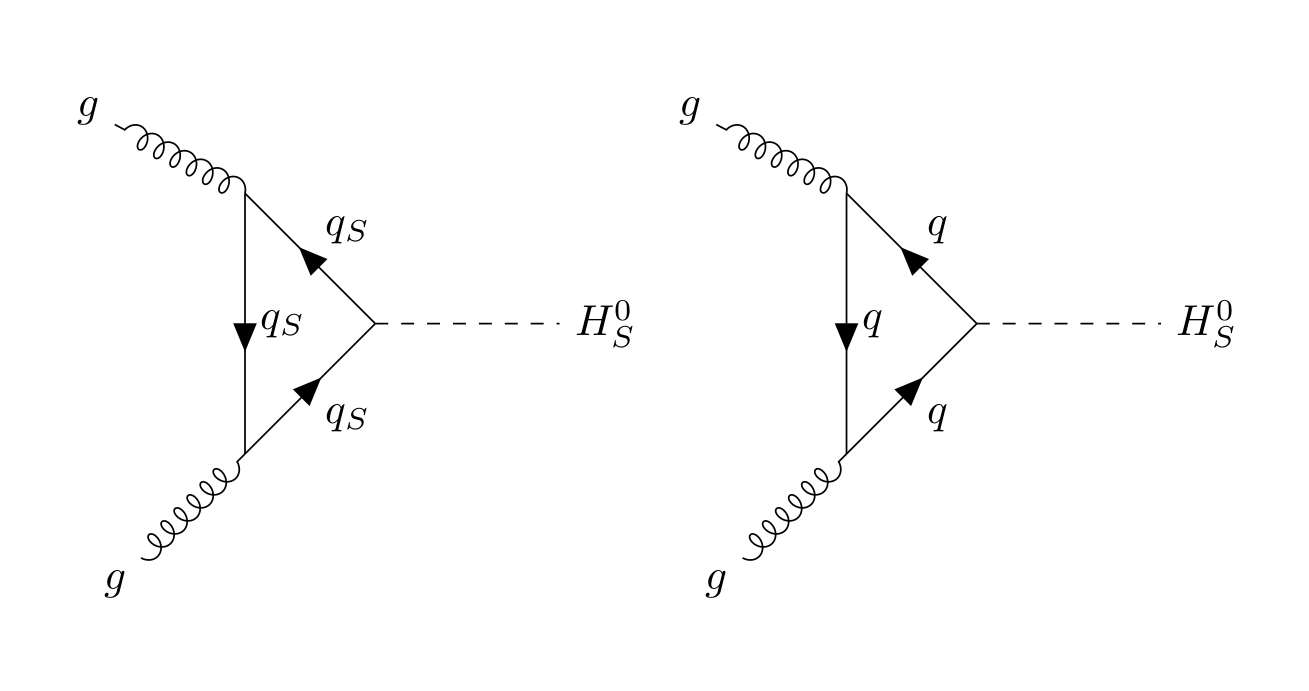}
		\end{center}
		\caption{Feynman diagram for $H_S^0$ production via gluon gluon fusion quark loops, $q_S$ is the exotic singlet quark and $q$ represents any SM quark}
		\label{singlet}
		\end{figure}

		 In this section we present singlet Higgs production cross-section via gluon gluon fusion over a range of singlet Higgs mass 
		 The production mechanism is the same as the SM Higgs production via gluon fusion. However,  the triangle loop (see Fig. \ref{singlet}) is driven by exotic quarks which are heavy in mass. There will be a very tiny contribution from the standard model top-quark
		 through the mixing of singlet Higgs with the SM Higgs boson. While estimating this  cross-section we have incorporated a K-factor following ref. \cite{qcdcor_ggh}, assuming higher order QCD correction to the production of a singlet Higgs boson will be of similar order that of a SM Higgs boson production via gluon fusion. For our illustration we have assumed $\beta_1$ (mixing parameter)
		 to be equal to 1.5 $\times 10^{-4}$. This value of $\beta_1$ is consistent with the measured values of SM Higgs boson signal strengths to different channels. 
		
	\begin{figure}[H]
	\centering
		\includegraphics[height=7.5cm, width=11cm]{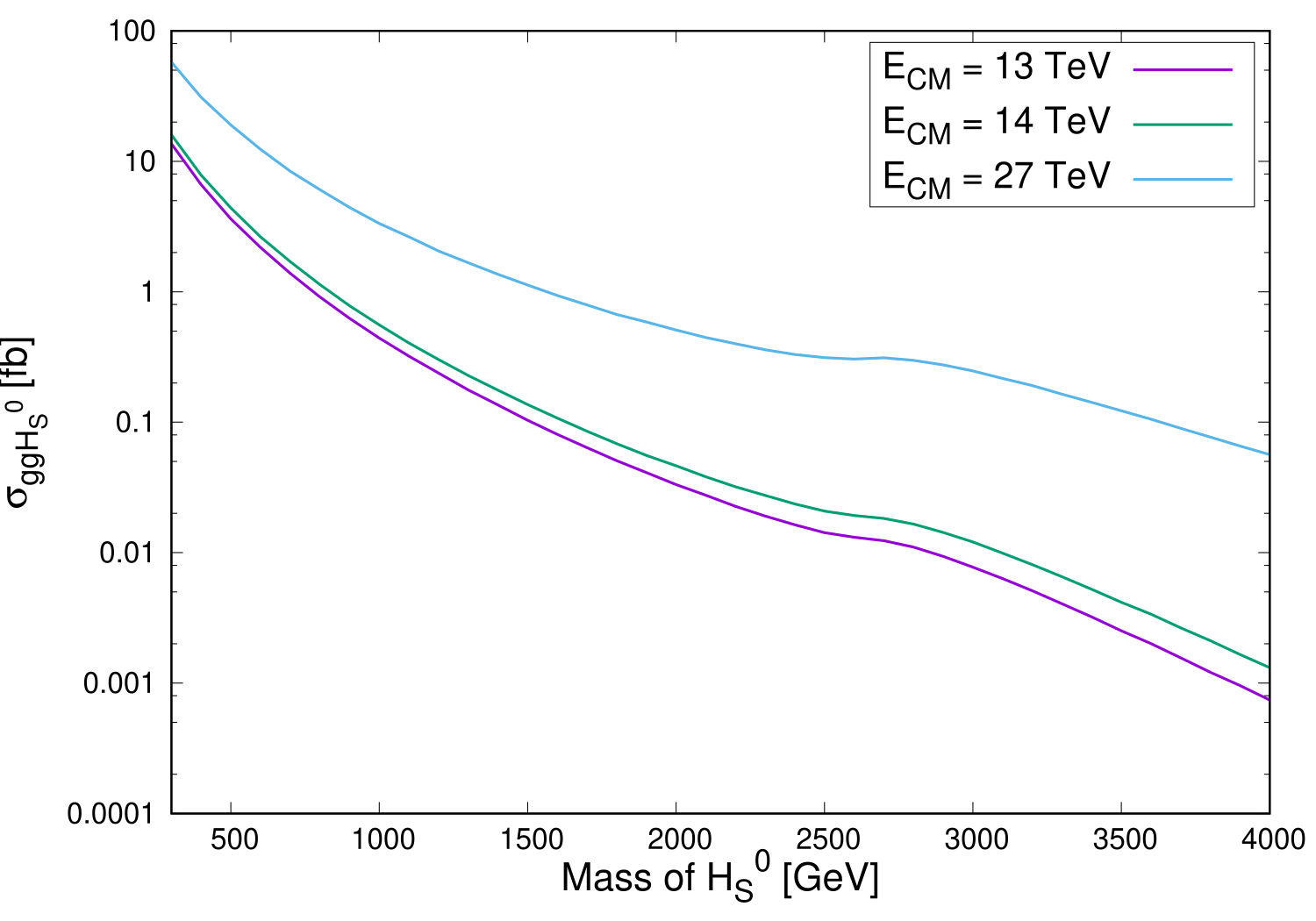}
		\scriptsize
		\caption{$H_S^0$ production cross-section ($\sigma$) via gluon fusion for  $\beta_1=1.5\times 10^{-4}$ and $v_S = 13$ TeV and 
		exotic quark mass 1.3 TeV.}
		\label{singlet_prod}
	\end{figure}

	In Fig. \ref{singlet_prod}, we have presented the singlet Higgs boson production cross-section at proton proton centre of mass energies 13, 14 and 27 Tev respectively.  At 14 (27) TeV run of LHC, production cross-section varies from 15.9 (57.6) fb to 10$^{-3}$ (0.056) fb when $H_S^0$ mass changes from 0.3 to 4 TeV. Although the production mechanism is similar to the SM Higgs boson production via gluon fusion, the cross-section for $H_S^0$ production is order of magnitude smaller than a SM-like Higgs boson of same mass, even after considering the contribution from 3 species of $SU(2)$ singlet exotic quarks. This can be explained by the small Yukawa coupling of these exotic quarks to the Singlet Higgs boson (see Eq. (\ref{LRYukawa})). We have assumed the exotic quark mass to be equal to 1.3 TeV \cite{exotic_quark}\footnote{The lower limit on the mass of a heavy stable quark following the ref. \cite{exotic_quark} is 200 GeV, obtained from 8 TeV run of LHC. Due to non availability of any further updated analysis at 13 TeV, we have assumed that mass limit, on such an object, is in the ballpark of a TeV. The mass limit on heavy stable lepton \cite{exotic_leptons} is 1.09 TeV. Assuming the quarks will have a higher production cross-section at the LHC, we have assumed they must be heavier than the exotic leptons.}.
	Before we conclude, let us make some qualitative comments about the possible signatures of $H_S^0$ at the LHC. For a low mass ($<700$ GeV), $WW$ decay mode can be exploited to look for possible signatures of this Higgs boson. However, once, $m_{H_S^0}$ becomes greater than a TeV, gluon gluon decay of $H_S^0$ becomes dominant and detection of such a scalar will be difficult due to a possible large QCD background. However, for higher singlet Higgs masses ($>2.2$ TeV), it can decay to a pair of exotic leptons, thus will produce a unique signature of two charged tracks with their invariant mass peaking at the singlet Higgs mass. This signal will be background free and probably the best bet for detection of such a scalar boson. As for example, at the 14 TeV LHC, decay of a 2.2 TeV $H_S^0$ will approximately produce 48 events with a pair of charged tracks with 3 ab$^{-1}$ integrated luminosity. While we expect to have 15 such events for a 3 TeV $H_S^0$ at 14 TeV run with same luminosity. At 27 TeV run, the situation will improve drastically, we can expect to see 80 such events even for a 4 TeV singlet Higgs boson.

	Finally we would like to make a brief comment about the situation when $c_1 + c_3 \ne 0$ and $\gamma_1 \ne 0$. 
	Making $c_1 + c_3$ non-zero, would introduce mixing between right-handed neutral scalar with SM-like Higgs boson. However, we have to satisfy the experimentally observed signal strengths of $h^0$. This in turn limits the above mixing and the $H^0 _R$ would possibly have small decay channels to the SM fermions and SM gauge bosons. On the other hand a non-zero $\gamma_1$ would have more prominent effect on $H^0 _R$ - $H^0 _S$ phenomenology. A $\gamma_1$ induced mixing between $H^0 _R$ and $H^0 _S$ would lead to $H^0 _R$ decays to SM fermions along with exotic fermions when kinematically allowed. At the same time, both these states could be produced via gluon fusion. 
	
	The production of the BSM scalars in 32121 model and the possible backgrounds have been discussed in Table \ref{table3} very briefly.
	
	\begin{table}[H]
		\centering
		\begin{tabular}{|c|c|c|c|}
			\hline \hline
			Scalars & Production at LHC & Possible final state & Possible backgrounds \\[1.0ex]
			\hline \hline
			$h_2^0 (\xi_2^0)$ & $h_2^0 (\xi_2^0) b \bar{b}$ & $ b \bar{b} b \bar{b}l^+ l^- \nu_l \bar{\nu}_l$ & $t \bar{t} b \bar{b}, t \bar{t} h,$ DY + jets \\[1.0ex]
			\hline
			$H_1^\pm$ & $H_1^+ \bar{t} b$ & $ b \bar{b} b \bar{b}l^+ l^- \nu_l \bar{\nu}_l$ & $t \bar{t} b \bar{b}, t \bar{t} h,$ DY + jets \\[1.0ex]
			\hline
			$H_L^\pm$ & $H_L^+ H_L^-$ pair-production & $H_L^+ H_L^-$ & Stable heavy charged particle \\[1.0ex]
			& & & creating two charged tracks and \\[1.0ex]
			& & & possibly background free \\[1.0ex]
			\hline
			$-$ & $H_L^+ h_L^0 (\xi_L^0)$ & $H_L^+ h_L^0 (\xi_L^0)$ & Being stable and neutral $h_L^0 (\xi_L^0)$ \\[1.0ex]
			& & & will remain undectected and $H_L^+$ \\[1.0ex]
			& & &  will create one charged track \\[1.0ex]
			& & & and possibly background free \\[1.0ex]
			\hline
			$H_R^0$ & $H_R^0 A'$ & $H_L^+ H_L^- l^+ l^-$ & Two oppositely chagred tracks of \\[1.0ex]
			& & & heavy stable particles, invariant \\[1.0ex]
			& & & mass distribution of $H_L^+ H_L^-$ should \\[1.0ex]
			& & & peak at $M_{H_R^0}$, background free \\[1.0ex]
			\hline
			$H_S^0$ & $H_S^0$ via gluon fusion & $E^+ E^-$ & Stable heavy charged particle \\[1.0ex]
			& & & creating two charged tracks and \\[1.0ex]
			& & & possibly background free \\[1.0ex]
			\cline{3-4}
			& & $q_S \bar{q}_S$ & Stable heavy charged colored \\[1.0ex]
			& & & particles will hadronize \\
			\hline \hline
		\end{tabular}
		\caption{Significant production processes of BSM scalars of 32121 at the LHC and their possible backgrounds}
		\label{table3}
	\end{table}

		
	\section{Conclusions}\label{section4}
	To summarise, we have  investigated phenomenological implications of a LR symmetric model based on $E_6$ inspired gauge group $SU(3)_C \otimes SU(2)_L \otimes U(1)_L \otimes SU(2)_R \otimes U(1)_R$.  The later symmetry group can be a result of two step breaking of $E_6$. We have studied the phenomenology of the Higgs bosons, responsible for the symmetry breaking of 32121 gauge group down to the SM gauge group.
	The model is hallmarked by the presence of a complete family of \textbf{27}-plet of fermions belonged to the fundamental representation of $E_6$. Apart from these TeV scale  fermions,  a weak bi-doublet (under $SU(2)_L \times SU(2)_R$),  a right-handed Higgs doublet and a singlet is necessary for complete  symmetry breaking. The measured value of $W$-boson mass fixes one of the bi-doublet vacuum expectation values, which we identify with the SM Higgs vev. The experimental lower limits on the mass of right-handed charged gauge boson $W_R$ in turn constrain the  vacuum expectation value, $v_R$, of the neutral member of the right-handed Higgs doublet. $v_R$ comes out to be greater than 14 TeV. The second bi-doublet vev $k_2$ is set to zero to avoid a possible admixture of $W_R$ in the SM $W$ boson.  Appearance of an additional massive neutral gauge boson is a result of the breaking of an extra $U(1)$ symmetry. The experimental lower limit from the LHC, on the mass of such an extra $U(1)$ gauge boson puts a lower limit of 12.61 TeV on the vacuum expectation value $v_S$ of the singlet scalar boson. 

	All fermions present in the $27$ dimensional fundamental representation of $E_6$ are considered to be present in our model. We have written down the relevant dimension-4 Yukawa interactions of these fermions either with the bi-doublet scalar or the singlet scalar field, excepting the singlet lepton field $L_S$, for which we write a dimension-5 Yukawa term involving singlet scalar $\Phi_S$. 
	Furthermore,  using the LHC data on the search of heavy charged long-lived particles, we have put a lower limit of 1.09 TeV on the mass of heavy exotic charged lepton. 
		
	After discussing the symmetry breaking pattern in some details, we have mainly devoted ourselves on  the phenomenology of the scalars (CP-even, CP-odd and charged Higgs) present in this  model.  We investigated their decay modes, and possible production processes at the LHC. Without going into the details of signal background analysis we have discussed possible signatures of the Higgs bosons arising from this model.
		
	Apart from the SM-like Higgs boson $h^0$, two more neutral scalars ($h^0_2$ and $\xi^0_2$) of same mass and having similar couplings to the SM fermions will originate from the bi-doublet  after SSB. Lower limit on the masses of these scalars have been obtained from the LHC data and they must be heavier than 800 GeV.  Their dominant production mechanism at the LHC will be in association with a pair of $b$-quarks. Once produced they will mainly decay to a pair of $b$-quarks. The production cross-section  of such scalars via gluon fusion vary from $14~(77)$ fb to $0.2~(1.5)$ fb for $m_{h_2^0} = 800$ and $1500$ GeV respectively at 14 (27) TeV run of LHC.

	Three more neutral Higgs bosons arise after SSB from the left and right-handed doublets. Two of them have their origin in the left-handed doublet and one in the right-handed doublet.  The neutral Higgses originating from the left-handed doublet are stable and once produced in the collider they can only contribute to missing energy signature. These scalars can be good candidate for relic of the Universe. While the scalar which arise from right-handed doublet can be produced at the LHC in association with any of the  neutral  gauge boson ($Z$, $Z'$ or $A'$)  or via vector boson fusion process. The production cross-section varies from  $0.4$ fb to $0.22$ fb for a range of Higgs mass $400$ to $1500$ GeV. 
		
	Two charged Higgs bosons will be the hallmark  of the model. One of them,  $H_1^\pm$ comes from the bi-doublet and this particular charged state mainly couples to a $t$ and a $b$ quark via Yukawa interactions. A lower  limit of 720 GeV has been derived on its mass from the LHC data. The estimated cross-section for $H_1^\pm$  production in association with $tb$  varies from  $0.15~(1)$ pb and $0.005~(0.06)$ pb for $m_{H_1^\pm} = 720$ and $1500$ respectively GeV at 14 (27) TeV center of mass energy at the LHC. The rest of the charged states have origin in the left-handed doublet. They can be produced at the LHC in a mechanism similar to Drell-Yan. Once produced they will not decay. But being charged, they will leave their signature in the detector via an ionising track. In SUSY models, similar signal are produced by stable/long-lived stau. Such a signal has been looked for at the LHC by ATLAS collaboration. A lower limit on the mass of the charged Higgs $H_L^\pm~( > 494)$ GeV has been derived using the ATLAS data. We further investigate the pair-production of $H_L^\pm$ and associated production of $H_L^\pm~H_L^0~(\xi_L^0)$ at LHC. 		
		
	The last menu in our list  is the singlet Higgs. It decays dominantly to a pair gluons and exotic fermions. We consider its production via gluon fusion at the LHC. However, production cross-section of a singlet Higgs via gluon fusion is inversely proportional to singlet vev, $v_S^2$. However, $v_S$ being in the ball park of 13 TeV, singlet Higgs production via gluon fusion fall below the level of a fb for a 1.5 TeV singlet Higgs even at 27 TeV run of the LHC.

Finally we would like to point out that, the Higgs sector of this model promises interesting phenomenology. A detail signal-background analysis has been already in our agenda \cite{32121_sig_bkg}. Finally, the neutral $SU(2)$ singlet lepton $N$, Higgs bosons $h^0_L$ and $\xi^0_L$ can serve the purpose of relic. It is important to see whether they can satisfactorily fulfil the constraints from the experimental data on relic density and direct detection of dark matter \cite{32121_dm}. \\
		
	{\bf Acknowledgement :} SB acknowledges financial support from DST-SERB, Govt.\;of India in form of an INSPIRE-Senior Research Fellowship. Both the authors are grateful to Prof.\;Joydeep Chakrabortty for introducing the subject to them and taking part in the initial stage 
	of the work. They also thank Dr.\;Triparno Bandyopadhyay for several insightful discussions and comments. SB also acknowledges Dr.\;Tapoja Jha and Debabrata Bhowmik for useful discussions. Both of us are grateful to Prof. Amitava Raychaudhuri for several discussions on the symmetries of scalar potential.		
		
		\section{Appendix:}
		
		\appendix
		
		\section{Masses and mixings in the particle sector of 32121 model:}
		
		\textbf{Neutral CP-even scalars:}
		
		\begin{eqnarray}
		\begin{pmatrix}
		h_1^0 \\ h_2^0 \\ h_L^0 \\ h_R^0 \\ h_S^0
		\end{pmatrix} = \begin{pmatrix}
		\cos\theta & 0 & 0 & 0 & \sin\theta \\
		0 & 1 & 0 & 0 & 0 \\
		0 & 0 & 1 & 0 & 0 \\
		0 & 0 & 0 & 1 & 0 \\
		-\sin\theta & 0 & 0 & 0 & \cos\theta
		\end{pmatrix} \begin{pmatrix}
		h^0 \\ h_2^0 \\ h_L^0 \\ H_R^0 \\ H_S^0
		\end{pmatrix}
		\end{eqnarray}
		where, \begin{equation}
		\theta = \dfrac{1}{2} \tan^{-1}\left(\dfrac{\beta_1 k_1 v_S}{\alpha_1 v_S^2 - \lambda_1 k_1^2}\right) \nonumber
		\end{equation}
		
		The basis on the LHS is the gauge eigenstates of the neutral CP-even scalars and the basis on the RHS shows the mass eigenstates, where $\theta$ is the mixing angle.

		\textbf{Charged scalars:}
		
		\begin{eqnarray}
		\begin{pmatrix}
		h_1^+ \\ h_2^+ \\ h_L^+ \\ h_R^+ 
		\end{pmatrix} = \begin{pmatrix}
		c_{11} & 0 & 0 & c_{14} \\
		0 & 1 & 0 & 0 \\
		0 & 0 & 1 & 0 \\
		c_{41} & 0 & 0 & c_{44} 
		\end{pmatrix} \begin{pmatrix}
		H_1^+ \\ H_2^+ \\ H_L^+ \\ H_R^+
		\end{pmatrix}
		\end{eqnarray}
		where,\begin{equation}
		c_{11} = \sqrt{1-\dfrac{k_1^2}{v_R^2}} = c_{44}, c_{14} = \dfrac{k_1}{v_R} = -c_{41} \nonumber
		\end{equation}
		
		\textbf{Neutral gauge sector:}
		
		\begin{eqnarray}
		\begin{pmatrix}
		W_{3L} \\ W_{3R} \\ B_L \\ B_R 
		\end{pmatrix} = \begin{pmatrix}
		a_{11} & a_{12} & a_{13} & a_{14} \\
		a_{21} & a_{22} & a_{23} & a_{24} \\
		a_{31} & a_{32} & a_{33} & a_{34} \\
		a_{41} & a_{42} & a_{43} & a_{44} 
		\end{pmatrix} \begin{pmatrix}
		Z \\ Z' \\ A' \\ A
		\end{pmatrix}
		\end{eqnarray}
		where $A$ is the photon and $a_{ij}$ are the elements of the mixing matrix or the rotational matrix that rotates the gauge basis to mass basis.

		\begin{eqnarray}
		a_{11} = \cos\theta_W, a_{21} = \dfrac{-g' \sin\theta_W}{g_{2R}}, a_{31} = \dfrac{-g' \sin\theta_W}{g_{1L}}, a_{41} = \dfrac{-g' \sin\theta_W}{g_{1R}} \nonumber \\
		a_{14} = \sin\theta_W, a_{24} = \sin\theta_W, a_{34} = \dfrac{\sqrt{\cos 2\theta_W}}{\sqrt{2}}, a_{44} = \dfrac{\sqrt{\cos 2\theta_W}}{\sqrt{2}} \nonumber \\
		a_{12} = -1.643\times 10^{-4}, a_{22} = 0.704, a_{32} = -0.707, a_{42} = 5.457 \times 10^{-2}  \nonumber \\
		a_{13} = 2.255 \times 10^{-5}, a_{23} = -0.450, a_{33} = -0.386, a_{43} = 0.804  \nonumber
		\end{eqnarray}
		
\pagebreak		
		\section{Couplings of the BSM scalars in 32121 model:}
		
		\textbf{For $h_2^0/\xi_2^0$:}
		\begin{table}[H]
			\centering
			\begin{tabular}{|c|c||c|c|}
				\hline \hline
				$M_{h_2^0}$ (GeV) & $\dfrac{1}{\sqrt{2}}\sqrt{4\lambda_3 k_1^2 + (c_4-c_3) v_R^2}$ & $M_{\xi_2^0}$ (GeV) & $\dfrac{1}{\sqrt{2}}\sqrt{4\lambda_3 k_1^2 + (c_4-c_3) v_R^2}$ \\[1.0ex]
				\hline
				$h_2^0 b \bar{b}$ coupling & $\dfrac{y_t}{\sqrt{2}}$ & $\xi_2^0 b \bar{b}$ & $\dfrac{y_t}{\sqrt{2}} \gamma^5$ \\[1.0ex]
				\hline
				$h_2^0 H^\pm W^{\mp}$ & $\dfrac{g_{2L}}{2} \sqrt{1-\dfrac{k_1^2}{v_R^2}}$ & $\xi_2^0 H^\pm W^{\mp}$ & $\dfrac{g_{2L}}{2} \sqrt{1-\dfrac{k_1^2}{v_R^2}}$ \\[1.0ex]
				\hline
				$h_2^0 t \bar{t}$ & $\dfrac{y_b}{\sqrt{2}}$ & $\xi_2^0 t \bar{t}$ & $\dfrac{y_b}{\sqrt{2}} \gamma^5$ \\[1.0ex]
				\hline
				\hline \hline
				\end{tabular}
				\end{table}

		\textbf{For $H_R^0$:}
		\begin{table}[H]
			\centering
			\begin{tabular}{|c|c|}
				\hline \hline
				$M_{H_R^0}$ (GeV) & $\sqrt{2\rho_1 v_R^2}$  \\[1.0ex]
				\hline
				$H_R^0 h^0 h^0$ & $- (c_1 + c_4) \cos^2\theta v_R - \gamma_1 \sin^2\theta v_R$  \\[1.0ex]
				\hline
				$H_R^0 H_L^\pm H_L^\mp$ & $- \rho_3 v_R$  \\[1.0ex]
				\hline
				$H_R^0 h_L^0 h_L^0/H_R^0 \xi_L^0 \xi_L^0$ & $- \rho_3 v_R$ \\[1.0ex]
				\hline
				$H_R^0 h^0 H_S^0$ & $(\gamma_1 - c_1 - c_4) \sin\theta \cos\theta v_R$ \\[1.0ex]
				\hline
				\hline \hline
			\end{tabular}
		\end{table}
		
		\textbf{For $H_S^0$:}
		\begin{table}[H]
			\centering
			\begin{tabular}{|c|c|}
				\hline \hline
				$M_{H_S^0}$ (GeV) & $\sqrt{2\alpha_1 v_S^2}$  \\[1.0ex]
				\hline
				$H_S^0 Z Z$ & $[\dfrac{g_{2L}^2 \cos^2\theta_W k_1}{2} + g_{2L} g' \sin\theta_W \cos\theta_W k_1 + \dfrac{g'^2 \sin^2\theta_W k_1}{2}] \sin\theta$  \\[1.0ex]
				\hline
				$H_S^0 h^0 h^0$ & $-\beta_1 \sin^3\theta k_1 + 2 (\beta_1 - 3 \lambda_1) \sin\theta \cos^2\theta k_1$ \\[1.0ex]
				&  $+ 2 (\beta_1 - 3 \alpha_1) \sin^2\theta \cos\theta v_S - \cos^3\theta \beta_1 v_S$ \\[1.0ex]
				\hline
				$H_S^0 W^\pm W^\mp$ & $\dfrac{g_{2L}^2 \sin\theta k_1}{2}$ \\[1.0ex]
				\hline
				$H_S^0 t \bar{t}$ & $\dfrac{y_t \sin\theta}{\sqrt{2}}$ \\[1.0ex]
				\hline
				$H_S^0 E^\pm E^\mp(H_S^0 N N)$ & $\sqrt{2} y_{LB} \cos\theta$ \\[1.0ex]
				\hline
				$H_S^0 q_S \bar{q}_S$ & $\dfrac{y_s \cos\theta}{\sqrt{2}}$ \\
				\hline \hline
			\end{tabular}
		\end{table}
		
		\textbf{For $H_1^\pm$:}
		\begin{table}[H]
			\centering
			\begin{tabular}{|c|c|}
				\hline \hline
				$M_{H_1^\pm}$ (GeV) & $\dfrac{1}{\sqrt{2}}\sqrt{(c_4-c_3)(k_1^2 + v_R^2)}$ \\[1.0ex]
				\hline
				$H_1^+ \bar{t} b$ & $-(y_t + y_b) \sqrt{1-\dfrac{k_1^2}{v_R^2}}$ \\[1.0ex]
				\hline 
				$H_1^+ h_2^0 W^-/H_1^+ \xi_2^0 W^-$ & $\dfrac{g_{2L}}{2} \sqrt{1-\dfrac{k_1^2}{v_R^2}}$ \\
				\hline \hline
			\end{tabular}
		\end{table}

	\end{document}